\documentclass[useAMS,usenatbib]{mn2e}
\usepackage{epsfig,graphicx,latexsym,amsmath,amssymb}
\usepackage{natbib}
\usepackage{hyperref}
\usepackage{mathrsfs}
\usepackage{wasysym}

\usepackage{latexsym}
\usepackage{graphicx}
\usepackage{eepic}
\usepackage{epic}
\usepackage{subfigure}
\usepackage{alltt}
\bibpunct[,]{(}{)}{;}{a}{}{,}

\title[Hybrid methods: Formation of protoplanetary systems]
       {Hybrid methods in planetesimal dynamics:\\
        Formation of protoplanetary systems and the mill condition}

\author[P. Amaro-Seoane, P. Glaschke \& R. Spurzem]
{
Pau Amaro-Seoane$^{1}$
\thanks{E-mail: Pau.Amaro-Seoane@aei.mpg.de (PAS);
                Glaschke@ari.uni-heidelberg.de (PG);
                Spurzem@ari.uni-heidelberg.de (RS)},
Patrick Glaschke$^{2}$ \&
Rainer Spurzem$^{3,\,4,\,2}$\\
$^{1}$Max Planck Institut f\"ur Gravitationsphysik
    (Albert-Einstein-Institut), D-14476 Potsdam, Germany\\
$^{2}$Astronomisches Rechen-Institut, M{\"o}nchhofstra{\ss}e 12-14,
      Zentrum f\"ur Astronomie, Universit\"at Heidelberg, Germany\\
$^{3}$National Astronomical Observatories of China, Chinese Academy of
      Sciences, 20A Datun Lu, Chaoyang District, 100012, Beijing, China\\
$^{4}$Kavli Institute for Astronomy and Astrophysics, Peking
      University, China
}

\begin{document}

\date{draft \today}

\pagerange{\pageref{firstpage}--\pageref{lastpage}} \pubyear{2013}

\maketitle

\label{firstpage}

\begin{abstract}
The formation and evolution of protoplanetary discs remains a challenge
from both a theoretical and numerical standpoint.
In this work we first perform a series of tests of our new hybrid algorithm presented in
Glaschke, Amaro-Seoane and Spurzem 2011 (henceforth Paper I) that combines the
advantages of high accuracy of direct-summation $N-$body methods with a
statistical description for the planetesimal disc based on Fokker-Planck
techniques. We then address the formation of planets, with a focus
on the formation of protoplanets out of planetesimals. We find that the evolution of the
system is driven by encounters as well as direct collisions and requires a
careful modelling of the evolution of the velocity dispersion and the size
distribution over a large range of sizes.
The simulations show no termination of the protoplanetary accretion due to gap
formation, since the distribution of the planetesimals is only subjected to
small fluctuations. We also show that these features are weakly correlated with
the positions of the protoplanets. The exploration of different impact
strengths indicates that fragmentation mainly controls the overall mass loss,
which is less pronounced during the early runaway growth.
We prove that the fragmentation in combination with the effective removal of collisional
fragments by gas drag sets an universal upper limit of the protoplanetary mass
as a function of the distance to the host star, which we refer to as the {\em mill condition}.
\end{abstract}


\section{Introduction\label{Introduction}}

The origin of our solar system remains to be one of the most exciting problems
of today's astronomy.  For a long time it has been the only known planetary
system.
While it is still the only planetary system that can be
studied in detail, progress in observation techniques has led to the discovery of
extrasolar planets and even some extrasolar planetary systems. The wealth
of observational data raised the question of how a planetary system forms in
general.
As of writing these lines, 859 planets and 676 planetary systems
are known\footnote{\url{http://exoplanet.eu/catalog-all.php}}. Most of these
planetary systems are very different compared to our solar system.

Understanding planet formation comprises many challenges, such as hydrodynamics
of the protoplanetary disc, chemical evolution of the embedded dust grains,
migration of planets and planetesimals and even star-star interactions in dense
young star clusters \citep[see][for a review and references therein, and also
the introduction of Paper I, for a brief summary]{Armitage2010}. All these
components constitute the frame for the essential process of planet formation:
An enormous growth from dust-sized particles to the final planets, accompanied
by a steady decrease of the number of particles which contain most of the mass
over many orders of magnitude. The particle number changes over many orders of
magnitude as planetary growth proceeds. There is active research on each of the
different aspects of planet formation, but the current efforts are far from a
unified model of planet formation \citep{Goldreich2004b,Lissauer1993}.

We address the study of this many--to--few transition from planetesimals to few
protoplanets. This stage is of particular interest, as it links the early
planetesimal formation to the final planet formation.  Collisions still play a
major role in the evolution of the system, and the close interplay between the
change of the size distribution and the evolution of the random velocities
requires a careful treatment of the complete size range.

Small $N-$body simulations (i.e. with less than few $10^4$ particles) have been
useful in exploring the basic growth mechanisms at the price of a modified
timescale and an artificially reduced particle number
\citep[e.g.][]{Koku1995,Koku1996}.  Statistical codes explored the limit of
large particle numbers in the early phases and are now tentatively applied to
the full planet formation process.  An  efficient solution would be the
combination of these two approaches in one hybrid code to unify the advantages
of both methods.

In this work, we present a series of tests and first results of our new hybrid
code, which was presented in Paper I. As described in the first
paper, it combines the {\sc Nbody6} code (a descendant of the
widespread $N-$body family \citep[see][]{Aarseth1999,Spurzem1999,Aarseth2003}
with a new statistical code
which uses recent
works on the statistical description of planetesimal systems.  The new hybrid
code includes a consistent modelling of the velocity distribution and the mass
spectrum over the whole range of relevant sizes, which allows us to apply a
detailed collision model rather than the perfect-merger assumption used in previous
$N-$body simulations.
We then apply this new code to follow the formation of protoplanets out of 1--10 km
sized planetesimals. In section~\ref{sec.validation} we present a series of tests
that check for the rubustness of the code. In section~\ref{sec.ICs} we explain the
initial conditions we use for our numerical experiments, which we show in section~\ref{sec.simulations}.
In section~\ref{sec.mill} we derive a useful relation that allows us to
introduce an universal upper
limit of the protoplanetary mass as a function of the distance to the host star.
Finally, in section~\ref{sec.discussion} we discuss our progress and results and potential
future applications.

\section{Validating the Code}
\label{sec.validation}

The new hybrid code requires the implementation of rather different methods
within a single framework. We have here two possible sources of problems.
First, the method is new and therefore it must be carefully assessed with other
work; on the other hand, the implementation must also be checked meticulously,
since it combines two rather different approaches.  We hence present in this
section a number of tests to check all code components, namely the evolution of
the velocity dispersion, the accuracy of the solver of the coagulation
equation, the proper joining of statistical and $N-$body component and an
overall comparison of statistical, $N-$body and hybrid calculations.
Table~\ref{SimData} summarises the selected test runs with the respective
initial conditions.

\begin{table*}
\begin{center}
\begin{tabular}{|l|l|l|l|l|l|l|l|l|}  \hline
  No. & $\Sigma$ & $\Delta a$ & $N$ & $N_{\mathrm{rad}}$ & $e^2/h^2$ & $i^2/h^2$ & $m$ & Type \\ \hline
  T1a & $1.1251\times 10^{-6}$ & 0.02 & 1000 & -- & 0.04 & 0.01 & $1.41\times 10^{-10}$  & $N-$body \\ \hline
  T1b & $1.1251\times 10^{-6}$ & 0.02 &  500 & 10 & 0.04 & 0.01 & $1.41\times 10^{-10}$  & Hybrid    \\ \hline
  T1c & $1.1251\times 10^{-6}$ & 0.02 &  --  & 10 & 0.04 & 0.01 & $1.41\times 10^{-10}$  & Statistic \\ \hline \hline
  T2a & $0.5626\times 10^{-6}$ & 0.08 &  800 & -- & 4    & 1 & $5\times 10^{-10}$ & $N-$body  \\ \hline
      & $0.5626\times 10^{-6}$ &      &  200 & -- & 4    & 1 & $2\times 10^{-9}$ & \\ \hline
  T2b & $0.5626\times 10^{-6}$ & 0.08 &  --  & 10 & 4    & 1 & $5\times 10^{-10}$ & Hybrid  \\ \hline
      & $0.5626\times 10^{-6}$ &      &  200 & -- & 4    & 1 & $2\times 10^{-9}$ & \\ \hline \hline
  T3  & Safronov               & --   &  --  & -- & --   & --&  --               & Statistic  \\ \hline \hline
  T4a & $1.1251\times 10^{-6}$ & 0.02  & 10.000   & -- &  4  &  1  & $1.41\times 10^{-11}$  & $N-$body \\ \hline
  T4b & $1.1251\times 10^{-6}$ & 0.02  & --  & 10 & 4  &  1  & $1.41\times 10^{-11}$  & Hybrid  \\ \hline
  T4c & $1.1251\times 10^{-6}$ & 0.02  & --  & 10 &   4  &  1  & $1.41\times 10^{-11}$  & Statistic \\ \hline \hline
  T5  & $1.8789\times 10^{-6}$ & --    & --  & -- &  620 &  155  & $2.4 \times 10^{-15}$  & Statistic \\ \hline
\end{tabular}
\end{center}
\caption[Parameters of all test simulations]
{Parameters of all test simulations (and hence preceded by a ``T''). The transition mass in T4b is $m_{\mathrm{trans}}=3.1\times 10^{-10}$ \label{SimData}
Only simulations T3, T4a--T4c and T5 include collisions. All values use $M_c=G=r_0=1$.}
\end{table*}

\subsection{Energy Balance}

The first test run is dedicated to a careful check of the interplay between
statistical component and $N-$body component with respect to the evolution of
the velocity dispersion. We therefore exclude from this first test collisions
and accretion.

\begin{figure}
\resizebox{\hsize}{!}
          {\includegraphics[scale=1,clip]{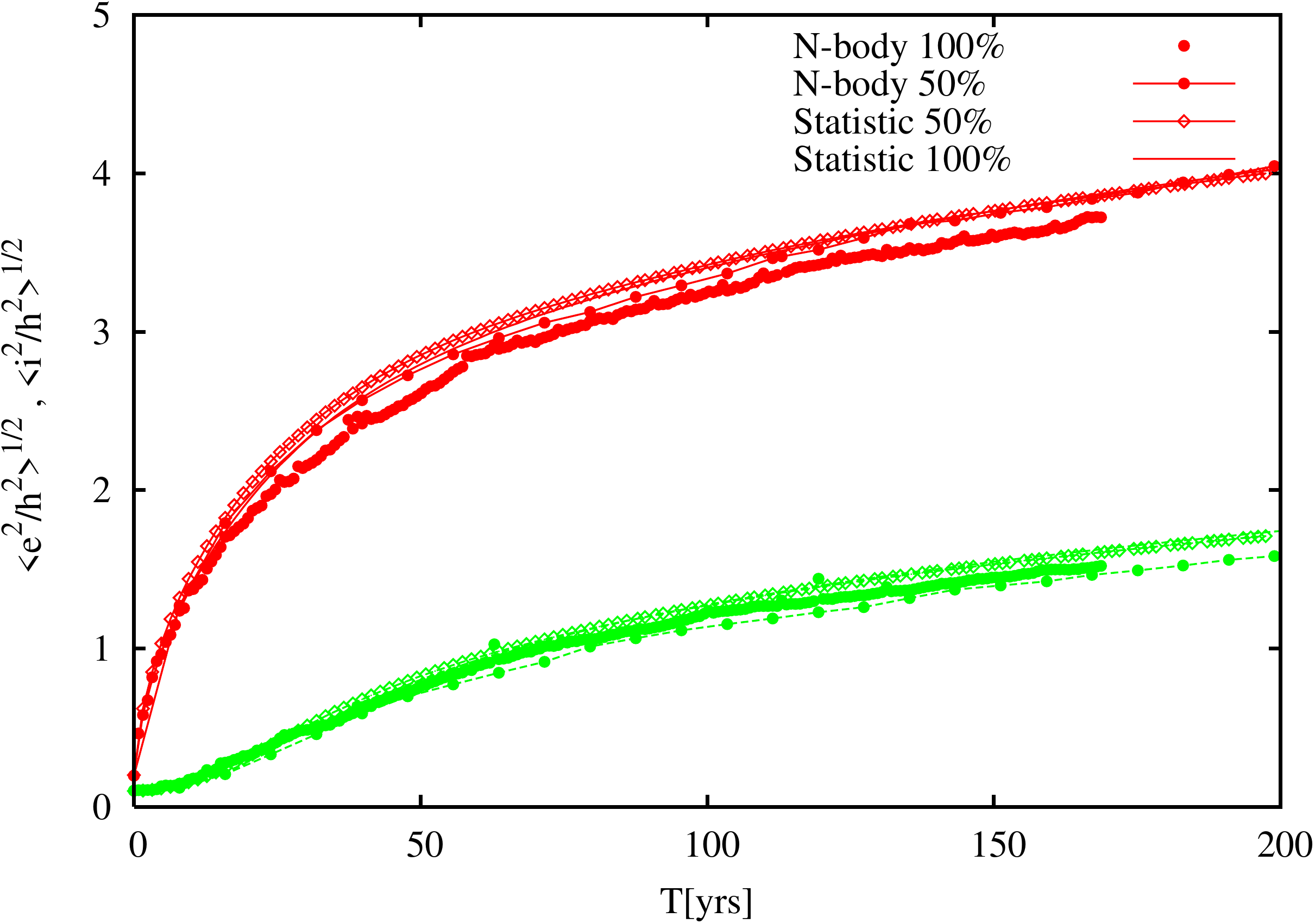}}
\caption
   {
Test simulations T1a--T1c (uniform mass, see table~\ref{SimData}). We show the results from
the $N-$body calculation (100\% $N-$body), the statistical calculation (100\% Statistic)
and the hybrid calculation (50\% Statistic refers to the statistical component,
whereas 50\% $N-$body is the $N-$body part). The red curve is the eccentricity data,
and the green curve the inclination.
   }
\label{DataLow}
\end{figure}

We use a homogeneous ring of planetesimals as our first test. The reason
is that we can analyse the evolution with three different setups -- a pure
$N-$body calculation, a pure statistical calculation and a mixed hybrid
calculation. All three approaches should in principle reproduce the same
result. Hence we prepare a small $N-$body test run (T1a) and let the system
evolve (see figure~\ref{DataLow}). As a second test run, we shift one half of
the bodies to the statistical model and conduct the integration again (T1b).
While this usage of the hybrid code is somewhat artificial, it provides an
excellent setup to examine the interplay between $N-$body and statistical part,
since neither component dominates the result. Finally, we run a complete
statistical calculation (T1c).

We can see that all different approaches are in good agreement. Although the
accordance between $N-$body and statistical calculation is not a new finding --
it merely shows that the stirring terms provide a proper description of a
planetesimal system (this was already shown by \cite{Ohtsuki2002}) -- we deem
the test to be necessary to demonstrate that the agreement holds in our
approach and, in particular, that the accuracy in the integration of the
statistical model is robust.  A more stringent test is posed by the hybrid run,
which proves that the pseudo-force method links both code components in a
consistent way without spurious energy transfer. In this respect,
figure~\ref{DataLow} includes both components of the hybrid calculation
separately, but the difference is so small that they are hardly
distinguishable.

\begin{figure}
\resizebox{\hsize}{!}
          {\includegraphics[scale=1,clip]{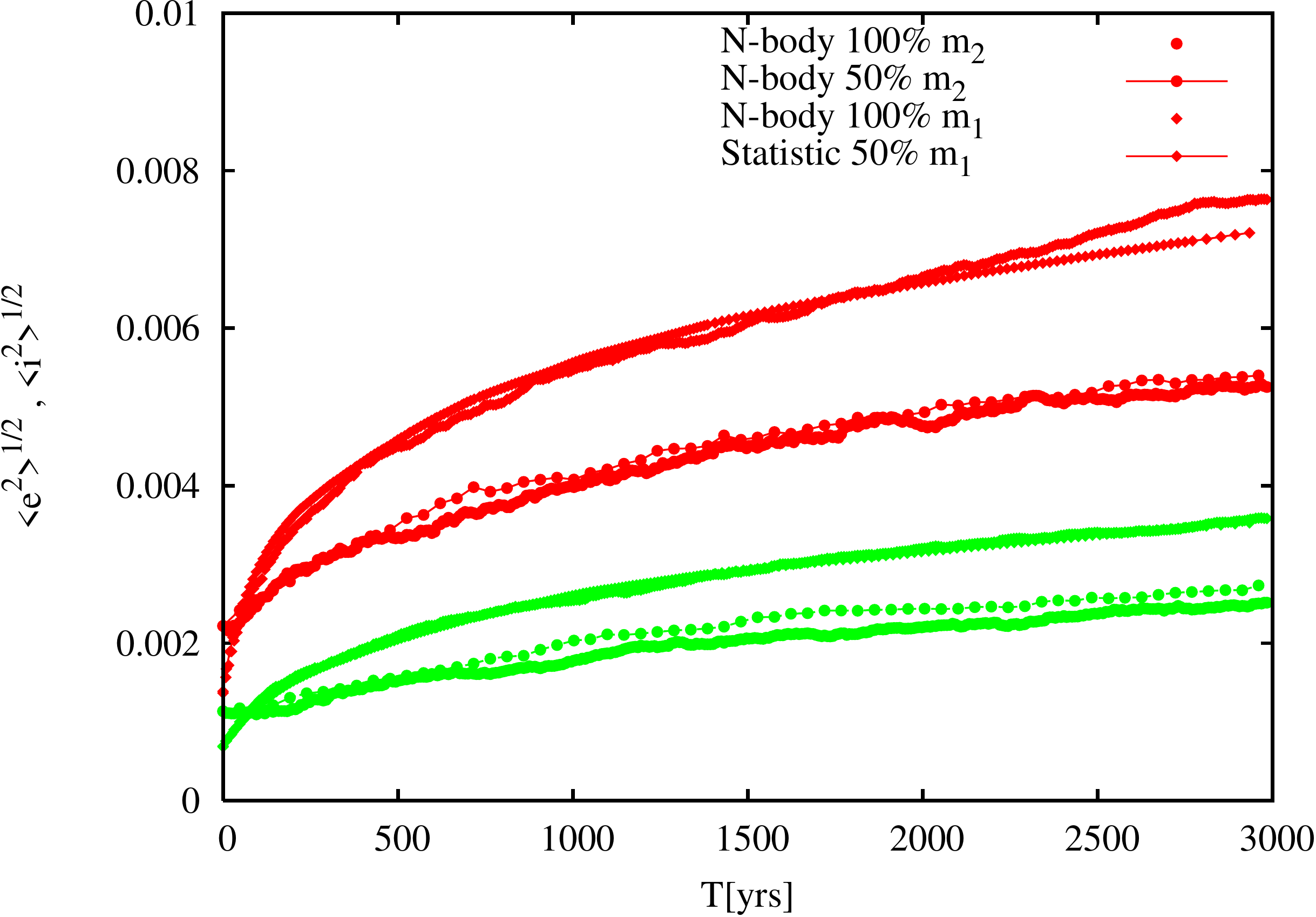}}
\caption
   {
Comparison of the $N-$body calculation T2a with the hybrid calculation T2b. The coding
is the same as in figure~\ref{DataLow}.
   }
\label{DataBimod}
\end{figure}

We also run a second test runs that follow the same approach but with a bimodal
mass distribution, with the same total mass in both components.  The first
case, T2a, is a pure $N-$body calculation, whereas the second one treats the
smaller particles with the statistical model. This test is particularly
interesting because it is close to the real purpose of our hybrid code. In
figure~\ref{DataBimod} we can see that there is a satisfactory agreement
between the two test runs.

\begin{figure}
\resizebox{\hsize}{!}
          {\includegraphics[scale=1,clip]{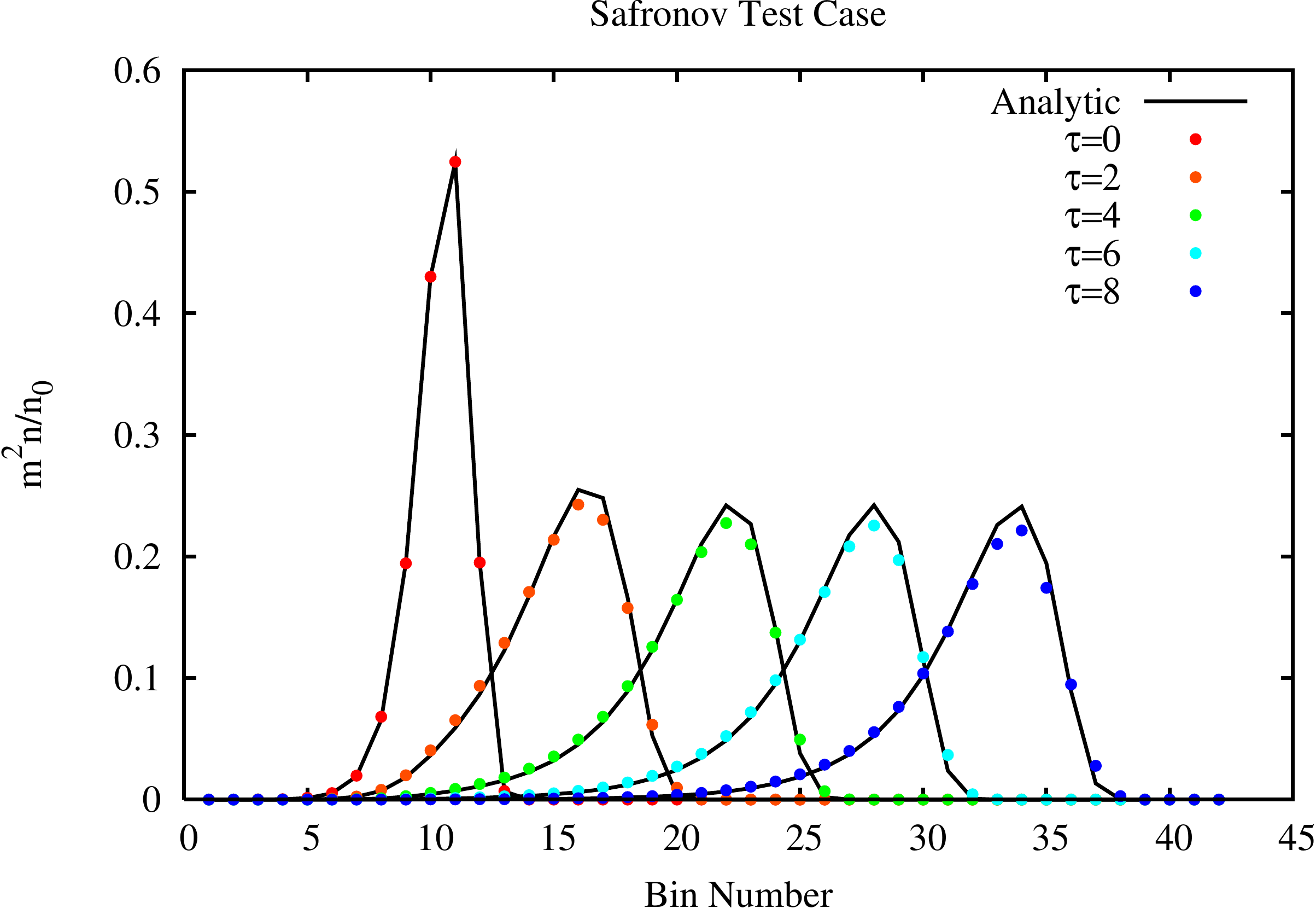}}
\caption
   {
Test of the solution of the coagulation equation (T3).
The analytical solution is presented in Paper I.
   }
\label{SafrTest}
\end{figure}

\subsection{Coagulation Equation}

In this section we verify the numerical solution of the coagulation equation by
running a comparison with the analytic solution of the Safronov problem, as
presented in Paper I.

The collisional cross-section is assumed to be proportional to the sum of the
masses of the colliding bodies. Thus, the coagulation kernel is known and an
additional integration of the velocity dispersions is not necessary.
Figure~\ref{SafrTest} summarises the numerical solution, simulation T3, of the
Safronov test.

The mass bins are spaced by a factor $\delta=2$. While some slight differences
emerge near the maximum of the density distribution, the overall shape is well
conserved throughout the integration. This proves that a spacing within a
factor two still guarantees a reliable solution of the coagulation equation
without a modified timescale for the growth.

\begin{figure*}
\resizebox{\hsize}{!}
          {\includegraphics[scale=1,angle=270,clip]{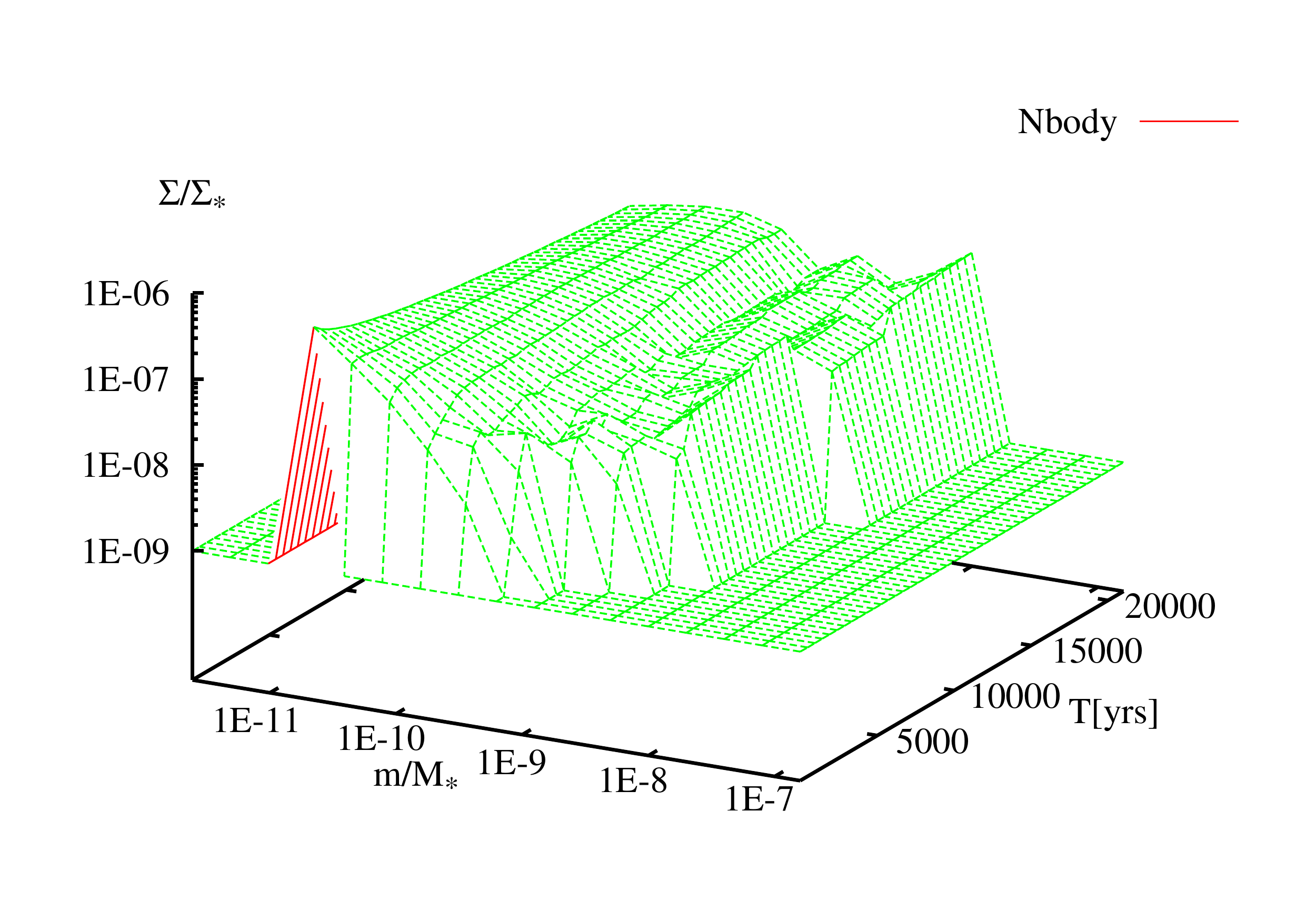}
           \includegraphics[scale=1,angle=270,clip]{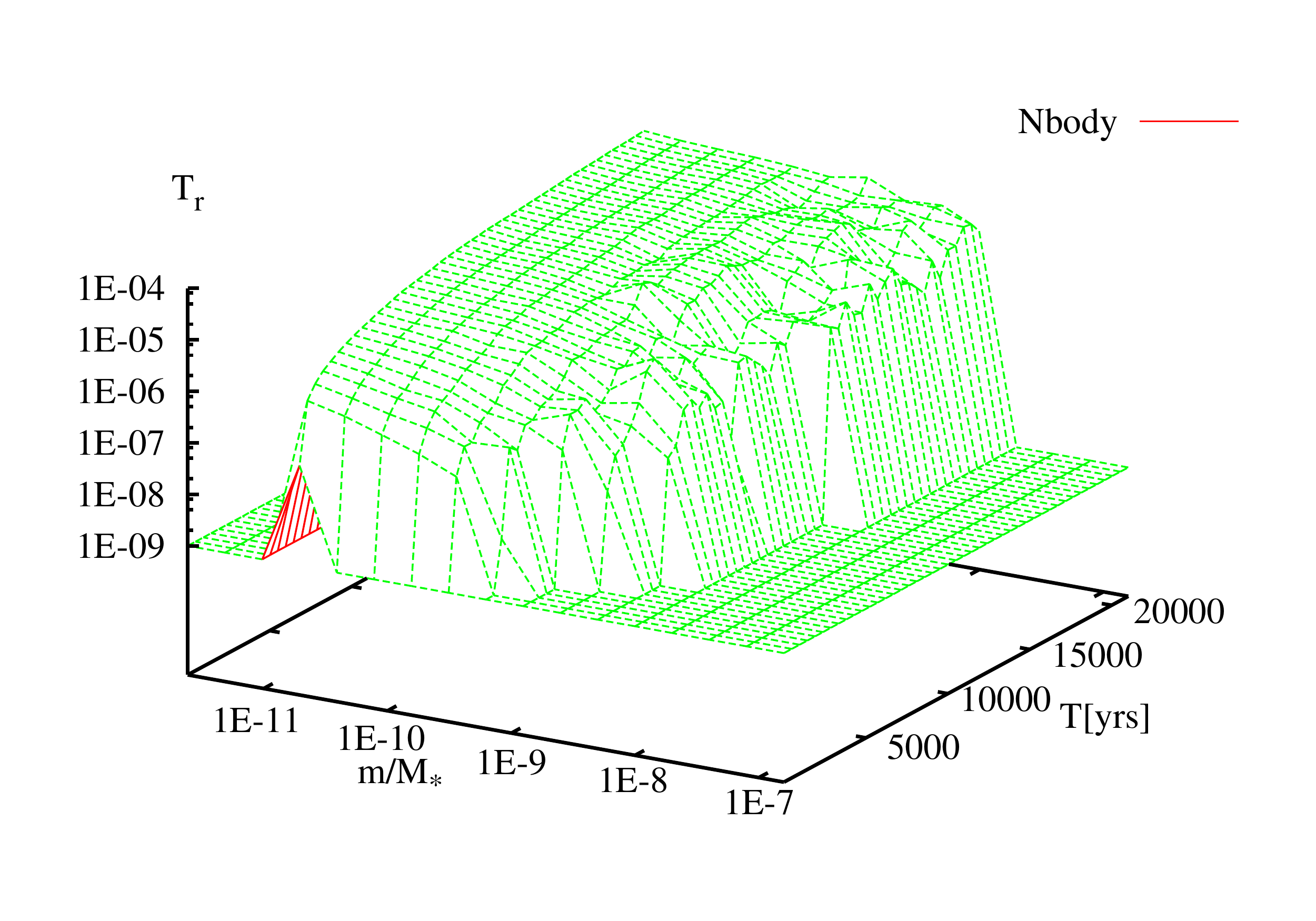}
          }
\caption
   {
Surface density and radial velocity dispersion of the $N-$body model (T4a).
   }
\label{L1}
\end{figure*}

\begin{figure*}
\resizebox{\hsize}{!}
          {\includegraphics[scale=1,angle=270,clip]{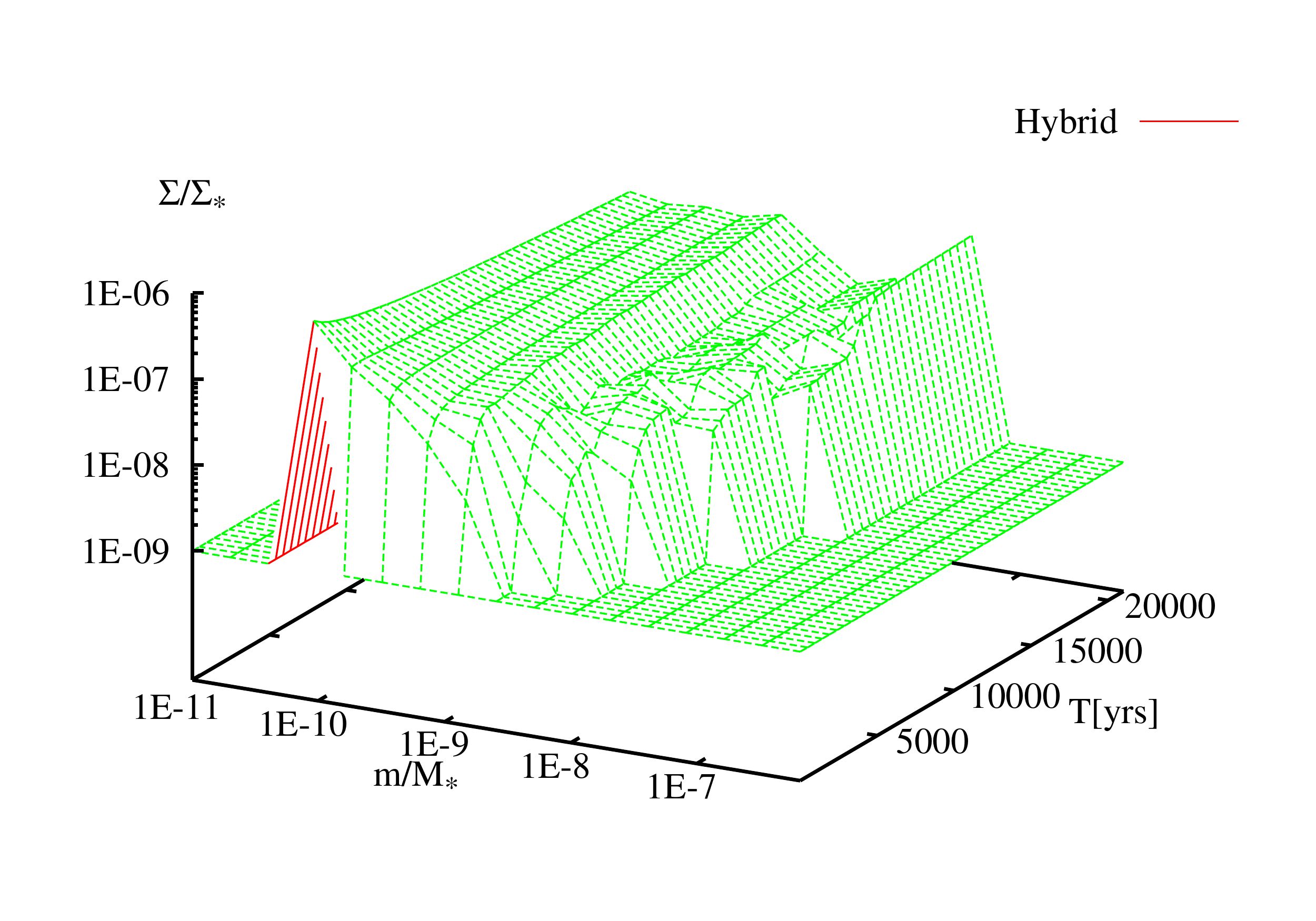}
           \includegraphics[scale=1,angle=270,clip]{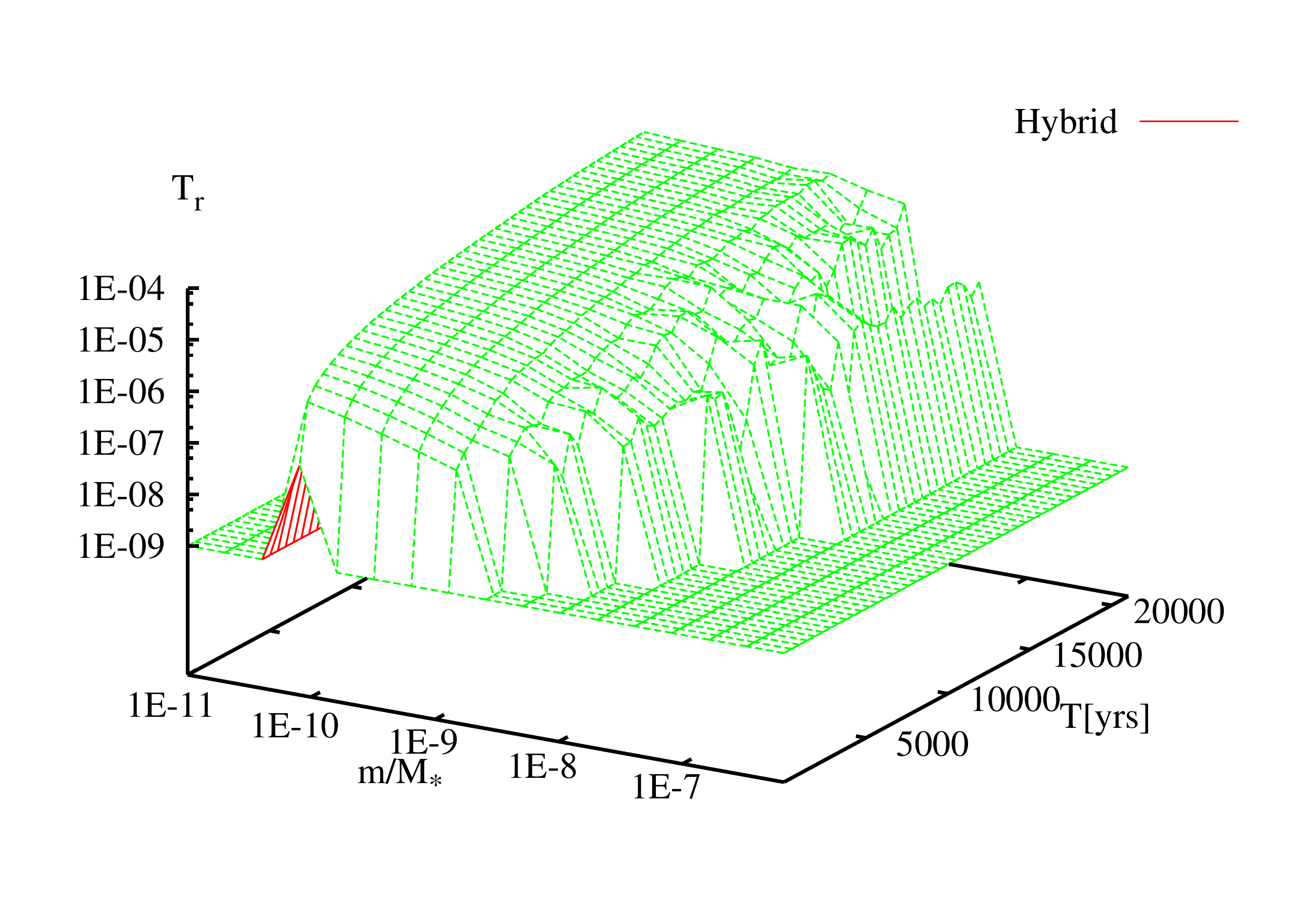}
          }
\caption
   {
Surface density and radial velocity dispersion of the hybrid model (T4b).
   }
\label{L2}
\end{figure*}

\begin{figure*}
\resizebox{\hsize}{!}
          {\includegraphics[scale=1,angle=270,clip]{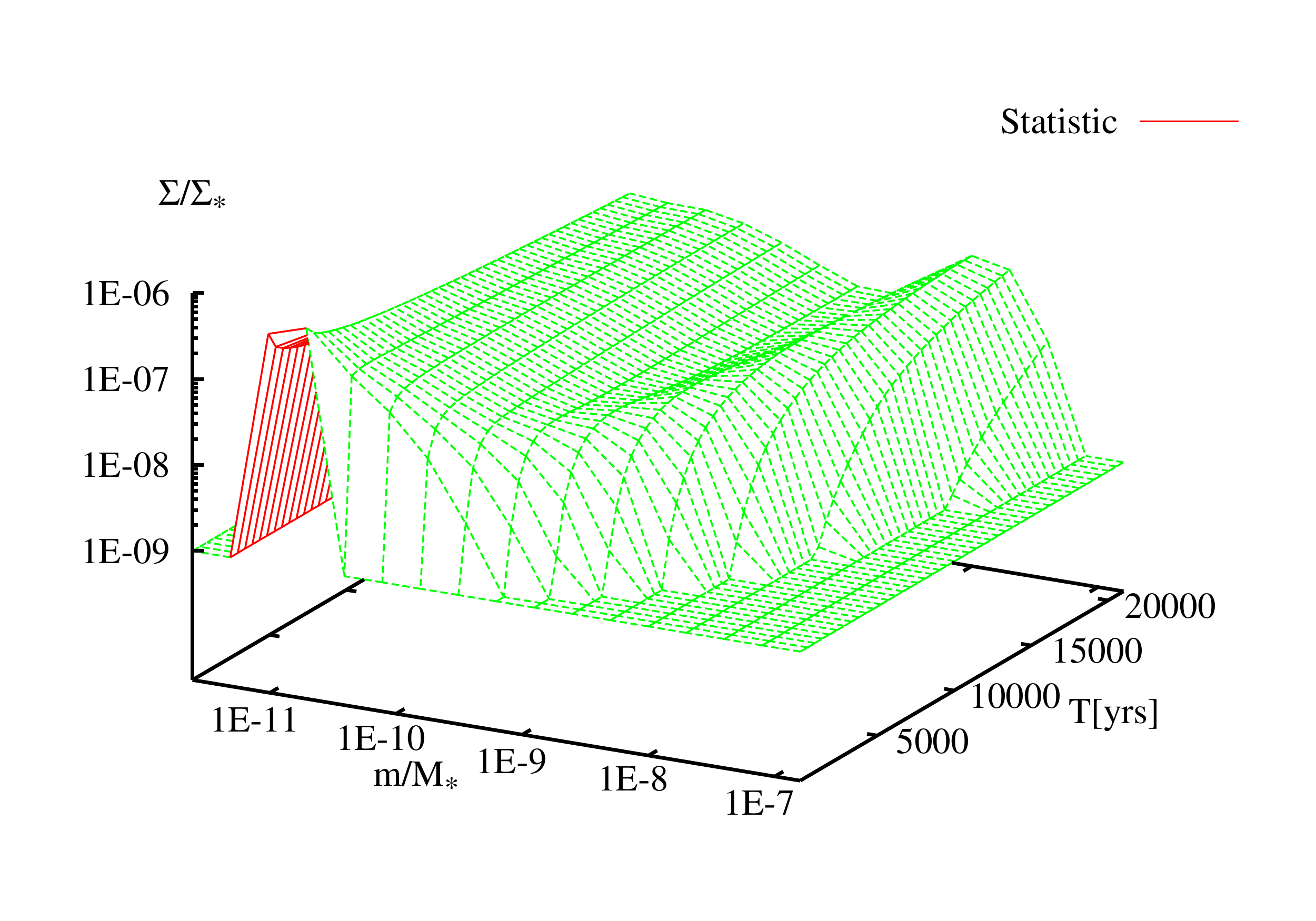}
           \includegraphics[scale=1,angle=270,clip]{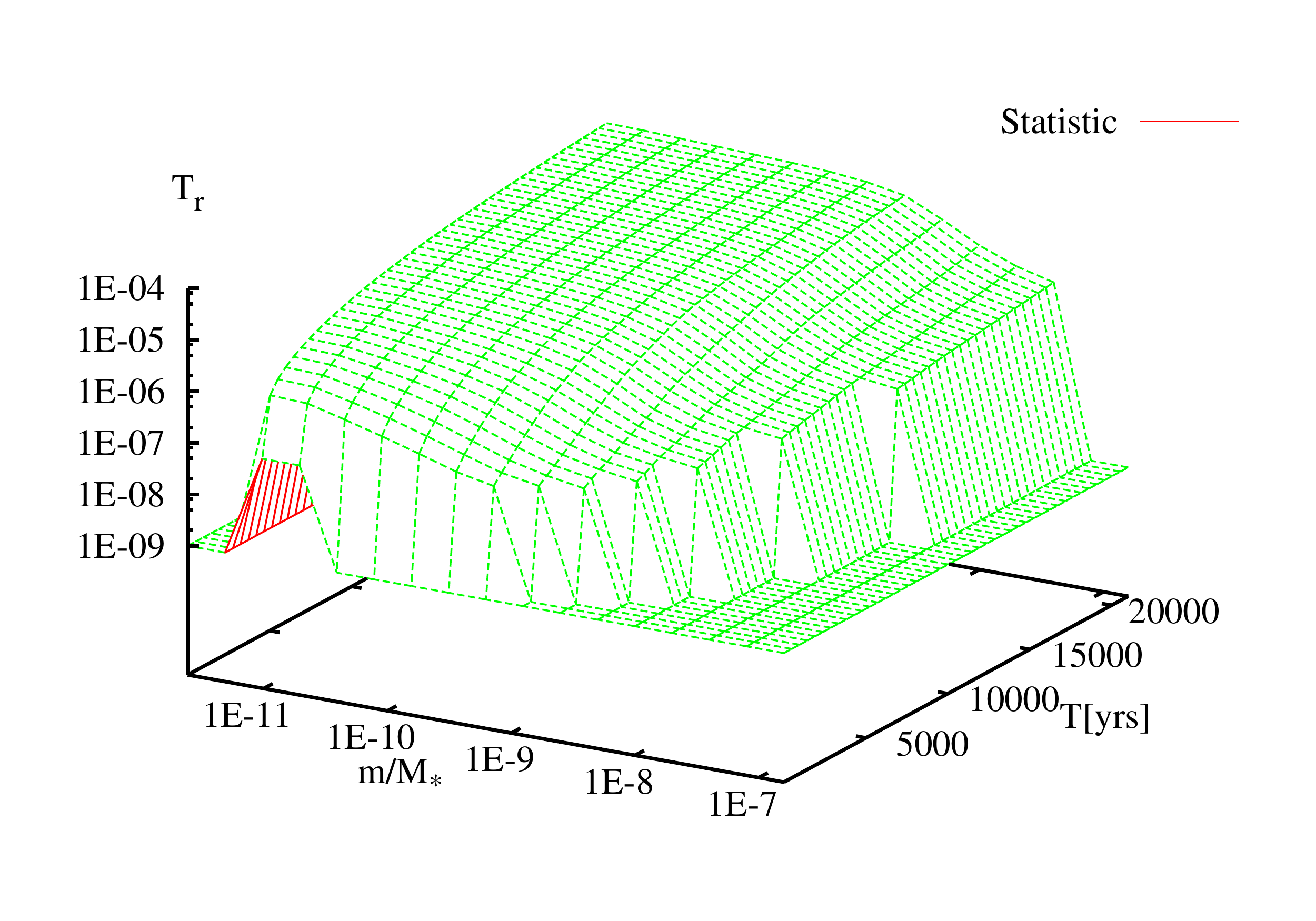}
          }
\caption
   {
Surface density and radial velocity dispersion of the statistical model (T4c).
   }
\label{L3}
\end{figure*}

\begin{figure}
\resizebox{\hsize}{!}
          {\includegraphics[scale=1,clip]{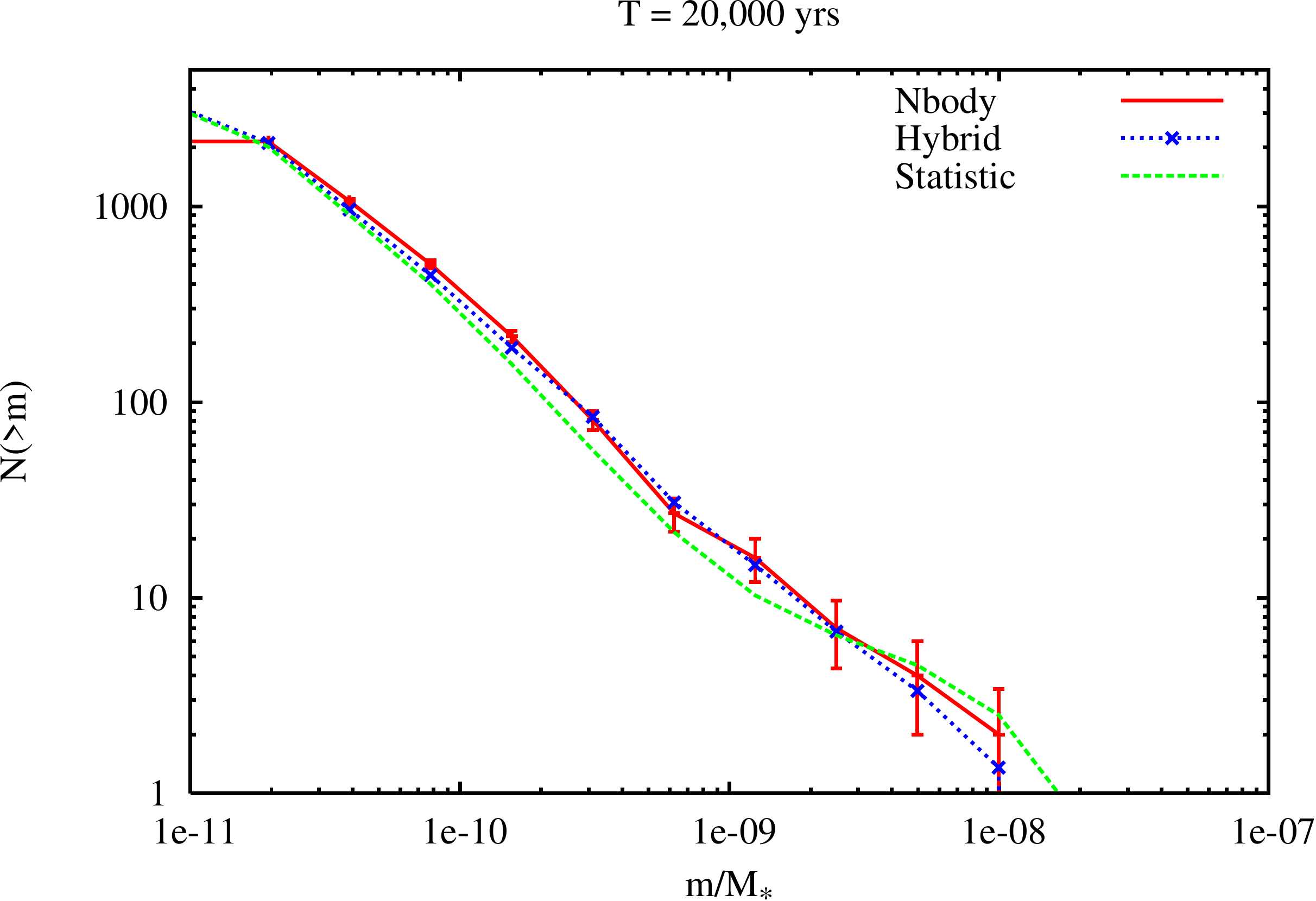}}
\caption
   {
Cumulative size distribution of the comparative runs T4a--T4c at T=20,000 yr.
   }
\label{xD1}
\end{figure}

\begin{figure}
\resizebox{\hsize}{!}
          {\includegraphics[scale=1,clip]{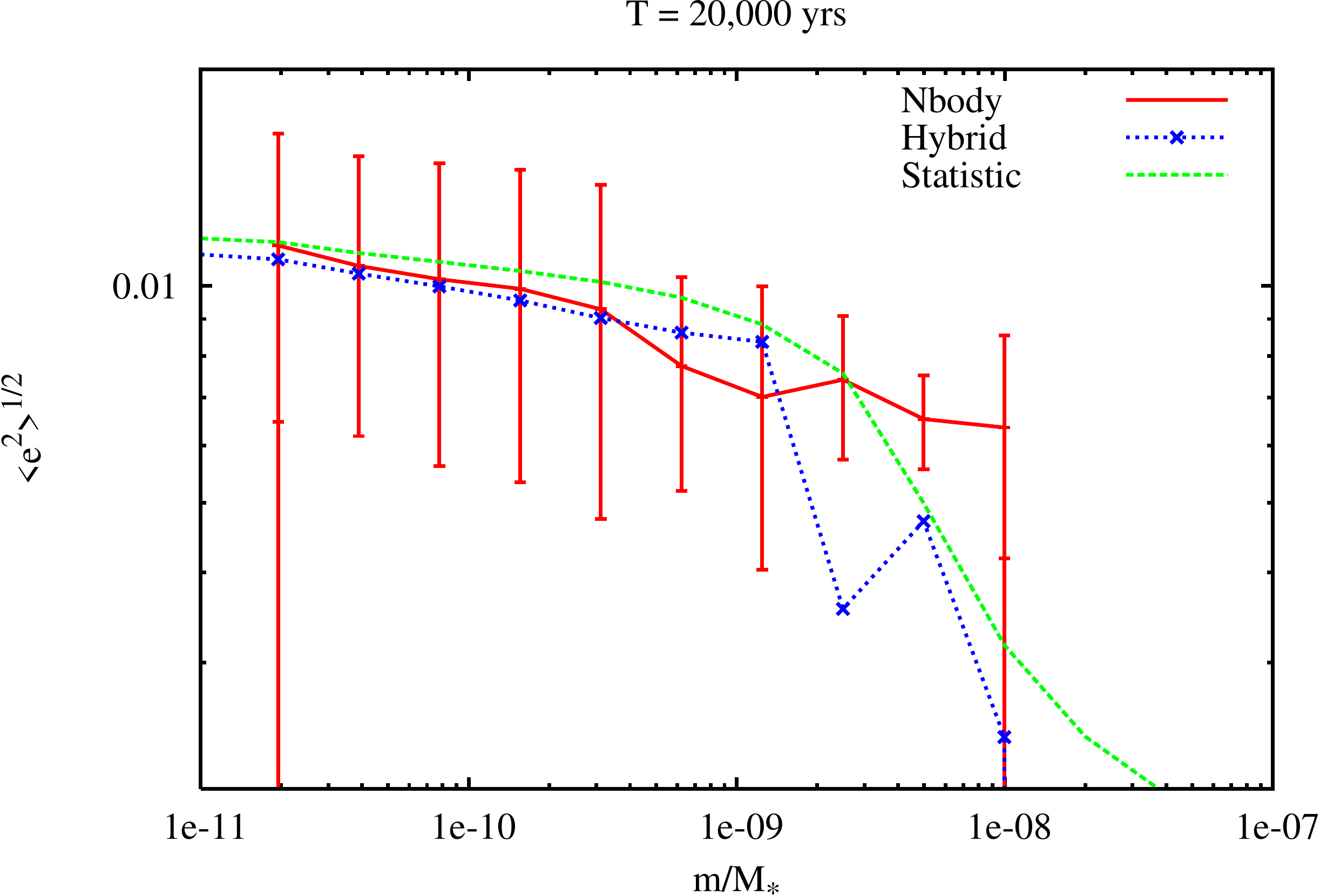}}
\caption
   {
Mean square eccentricities of the comparative runs T4a--T4c at T=20,000 yr. Error bars
indicate the spread due to the rayleigh distribution of the eccentricity.
   }
\label{xD2}
\end{figure}

\begin{figure}
\resizebox{\hsize}{!}
          {\includegraphics[scale=1,clip]{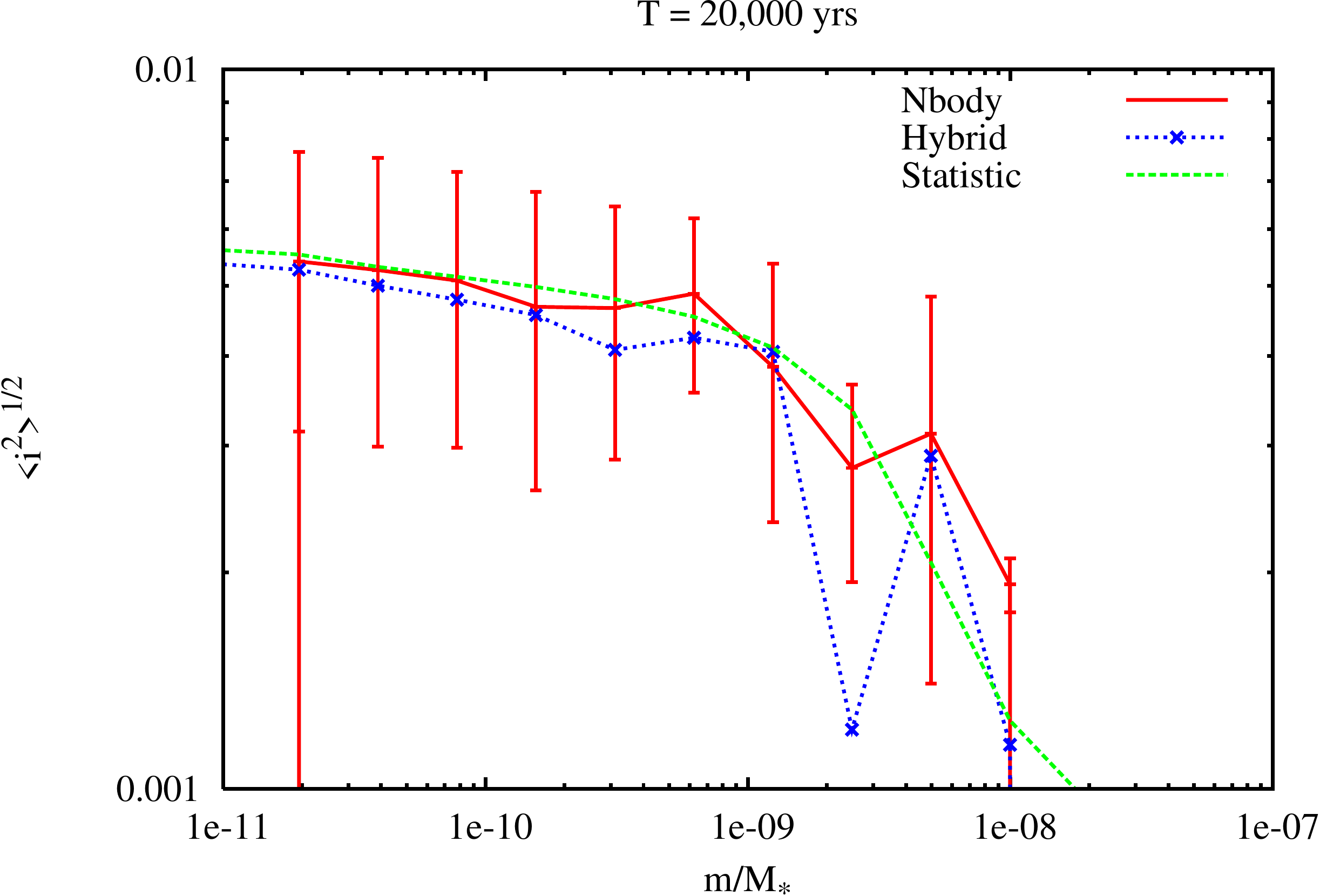}}
\caption
   {
The same as figure~\ref{xD2} for the inclination.
The strong deviation at $m=3\times 10^{-9}$ is due to a single particle.
   }
\label{xD3}
\end{figure}

\subsection{Testing the complete code}

The most robust test of our hybrid code (or the stand-alone statistical code)
is a comparison with a pure $N-$body simulation with the same initial
conditions. While a large particle number is desirable to cover a large range
in masses, we are limited in the number of particles to be used in the direct
$N-$body techniques to a few $10^4$.

\begin{figure}
\resizebox{\hsize}{!}
          {\includegraphics[scale=1,clip]{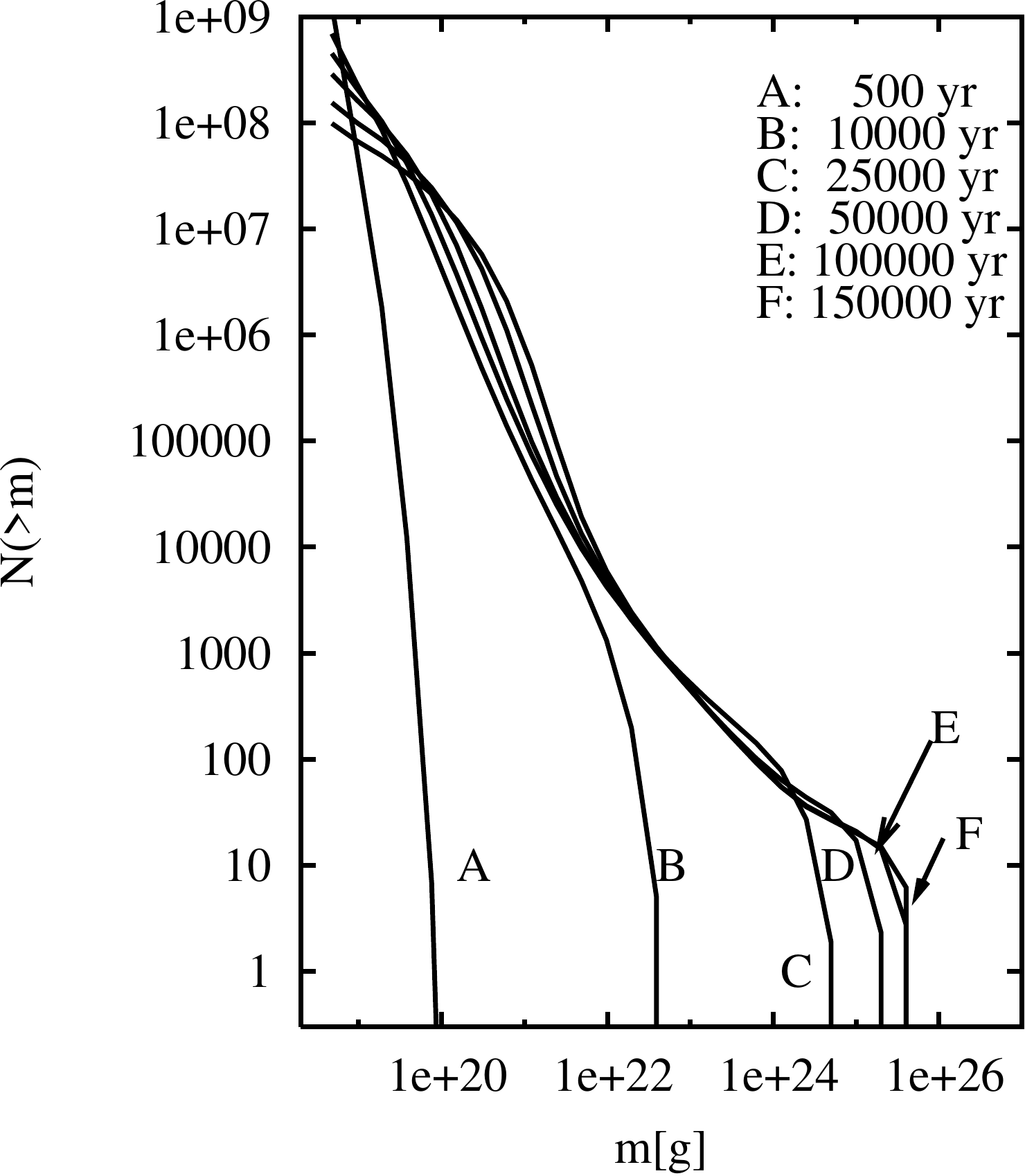}
           \includegraphics[scale=1,clip]{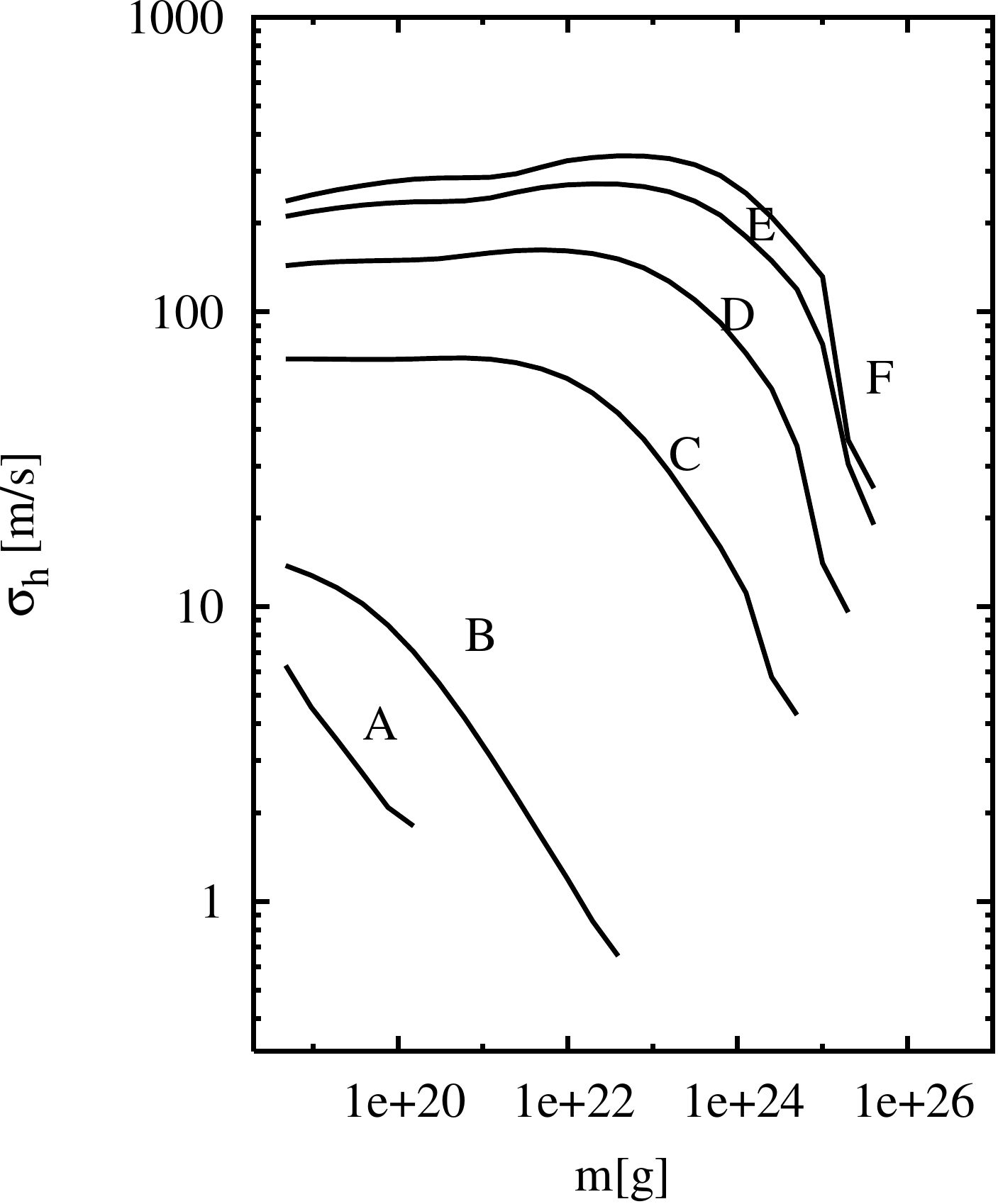}
          }
\caption
   {
Comparative  calculation T5 which adopts the initial conditions
of Inaba et al. 2001 (their figure 9, bottom).
   }
\label{VglIna}
\end{figure}

We therefore choose a single-mass system with initially 10,000 particles.  We
enlarge the radii of the planetesimal by a factor $f=5$, which speeds up the
calculation without modifying the growth mode. The transition mass is twenty
times larger than the initial planetesimal mass, keeping the particle number
covered by the statistical component larger than a few thousands.

We compare a full $N-$body run with a hybrid calculation and a pure statistical
calculation.  Though the stand-alone statistical calculation includes the
proper treatment of the runaway bodies via the gravitational range method, if
only few particles reside in one mass bin we do not take into account
suppression of self-accretion and self-stirring.  While the hybrid approach
describes this regime in much more detail, we include the full statistical
calculation nevertheless for completeness.

In figures~\ref{L1}--\ref{L3} we have an overview of the time evolution of the
system, where all quantities are integrated over the whole system.  All
calculations seem to agree rather well, although the statistical noise in the
$N-$body calculation and the hybrid calculation is quite strong due to the
particle number. Runaway growth leads to the fast formation of a few
protoplanets on a timescale of a few thousand years, with a good agreement of
the fast initial growth phase in all three test runs.  The boundary between
smooth evolution and noisy data marks the location of the transition mass in
the hybrid calculation.

We compare the size distribution and the velocity dispersion at the end of the
integration, which is 20,000 yr, in more detail in figures~\ref{xD1}--\ref{xD3}. Both the
$N-$body data and the hybrid data are projected on to the same grid as the full
statistical calculation to allow a convenient comparison.

The agreement of the size distribution $N(>m)$ is excellent; the small
deviations are within the statistical error.  We note that the strong
variations in the size distributions of figures~\ref{L1}--\ref{L3} are located
at the high mass end, where only few particles dominate the surface density. In
addition, the growth in the statistical model seems to be faster than the
$N-$body reference calculation.  However, the density at the highest masses
refers to less than one particle. As we noted before, this is due to the poor
treatment of the few--body limit.

The comparison of the velocity dispersions yields good results, in particular
in the low--mass regime, where the statistical error is small. The high mass
regime does not only suffer from bad statistics, but also from a pronounced
time variability, as we can see by comparing the fluctuations in
figures~\ref{L1}--\ref{L3}.  Taking these variations into account, all three
calculations are in good agreement.  As before, the deviation at $m=3\times
10^{-9}$ is due to a single particle.

\subsection{The statistical code}

\cite{Ina2001} presented a high accuracy statistical code. In this section we
run our last test calculation by comparing with this work, in particular with
their figure 9, bottom.  They included in their code approximately the same
physics and interpolation formulae, with only minor differences to our
approach.

While our approach allows us to set a spacing up to $\delta=2$, their solution
of the coagulation requires a smaller spacing, of $\delta=1.1$, to guarantee a
reliable solution.  The few-body limit is handled properly, with an additional
treatment of the protoplanets via the gravitational-range approach. In
figure~\ref{VglIna} we show our comparison,
simulation T5, with their runs. Again, we find a good agreement but for minor
deviations. These are likely related to the different implementation of the
collisional probability.

\section{Simulations: Protoplanetary growth}

\subsection{Initial conditions}
\label{sec.ICs}

We apply our hybrid code now to a well defined initial setup of a planetesimal
disc.  All simulations use a homogenous ring of planetesimals extending from an
inner boundary $R_{\min}$ to an outer boundary $R_{\max}$. Since radial
migration is not included, all planetesimals are bounded to this volume
throughout the simulation. The central star has a mass of one solar mass. Each
simulation starts with no $N-$body particles, so that we need only to specify
the setup for the statistical part of the calculation.  The differential
surface density as a function of mass is

\begin{align}
\frac{d\Sigma}{dm} &=  \Sigma_0 \frac{m}{{\bar m}^2}\exp\,\left(-m/\bar m\right)  \label{EqSigInit} \\
{\rm Var}(m) &=  {\bar m}^2,
\end{align}

\noindent
where $\Sigma_0$ is the total surface density and $\bar m$ is the mean mass.
Equation~\ref{EqSigInit} provides a smooth variation over a few mass bins,
which avoids numerical problems at the beginning of the simulation. The initial
velocity dispersion is related to the mean escape velocity $v_{\infty}$ of the
initial size distribution defined by {Eq.~\ref{EqSigInit}}

\begin{equation}
 \frac{1}{100} v_{\infty}^2  =   T_r +T_{\phi} +T_z,
\end{equation}

\noindent
with the ratio of the velocity dispersions

\begin{equation}
   T_r = 4T_{\phi} =4T_z
\end{equation}

We adopt a rather small initial velocity dispersion to avoid strong spurious
fragmentation due to an overestimation of the velocity dispersion. Furthermore,
strong relaxation in the initial phase of the calculation quickly establishes
an equilibrium velocity dispersion. The time step control parameters are chosen
such that the energy error $\Delta E/E$ of the $N-$body component remains
always smaller than  $10^{-8}$ throughout the simulation. Likewise, our choice
of the parameters of the statistical component assures that the statistical
model is solved accurately and remains stable, as indicated by the set of
comparative runs. All runs simulate only a narrow ring centred at a distance
$r_c$ and we choose the following units:

\begin{equation}
 r_c=1 \qquad M_c=1 \qquad G=1 \label{SimUnits}
\end{equation}

In table~\ref{ComParam} we summarise the main parameters of the simulations, fixed to
the same values for all of them.

\subsection{Main objectives of the analysis}
\label{sec.simulations}

The scheme we have developed is in principle ready to solve the complete
planetesimal problem, at least concerning the large range of sizes. However, in
practise we are limited by the computational power. A small ring with a width
of 0.1 AU centred at 1 AU with a moderate size for the lower cut-off requires
some days of integration, with the largest fraction of time spent in the
statistical model. While we focus on these initial conditions for our
simulations, we also present some more refined models that required larger
calculations.  We adapt a surface density $\Sigma=10$ g/cm$^2$ in the
simulations, which can be envisaged as a nominal value used in the related
literature. In the remaining of this work, we focus on the following aspects of
protoplanetary growth:

\begin{table}
\begin{center}
\begin{tabular}{|l|rl|} \hline
$\eta_{\mathrm{Disc}}$     & 0.01  &\\  \hline
$\eta_{\mathrm{reg}}$   & 0.002 &\\  \hline
$\eta_{\mathrm{irr}}$   & 0.001 &\\  \hline
$R_{\min}$ & 0.95 & AU \\   \hline
$R_{\max}$ & 1.05 & AU \\   \hline
$\bar m$ & $3\times 10^{18}$ & g \\ \hline
$\rho$     & 2.7  & g/cm$^3$ \\   \hline
$\delta$     & 2   &  \\   \hline
$\Delta v_g$ & 60 & m/s \\        \hline
\end{tabular}
\end{center}
\caption[General simulation parameters]{General parameters common to all simulations listed in table~\ref{ComDetail}.  \label{ComParam}}
\end{table}
\begin{table*}
\begin{center}

\begin{tabular}{|l|l|r|r|r|r|r|r|r|r|} \hline
Code   & Strength & $N_M$ & $N_R$ & $\Sigma[$g$/$cm$^2]$  & $m_{\min}/M_c$ & $m_{\mathrm{trans}}/M_c$ &$\rho_g[$g$/$cm$^3]$  \\ \hline
S1FB   & B\&A 1999      & 24 & 50  & 10  & $ 3.48 \times 10^{-18}$& $ 3.89 \times 10^{-11}$ & $10^{-9}$  \\ % FRAG
S2FH   & H\&H 1990      & 24 & 50  & 10  & $ 3.48 \times 10^{-18}$& $ 3.89 \times 10^{-11}$ & $10^{-9}$  \\ % FRAG\_HOUSEN
S3FN   & Perfect Merger & 24 & 50  & 10  & $ 3.48 \times 10^{-18}$& $ 3.89 \times 10^{-11}$ & $10^{-9}$  \\  % NOFRAG
S4FBN  & B\&A 1999     & 24 & 50  & 10  & $ 3.48 \times 10^{-18}$& $ 3.89 \times 10^{-11}$ & $ 0 $      \\ \hline% FRAG\_NOGAS
S5FBL  & B\&A 1999      & 24 & 5   & 10  & $ 3.48 \times 10^{-18}$& $ 3.89 \times 10^{-11}$ & $10^{-9}$  \\ % FRAG\_low
S6FBH  & B\&A 1999      & 24 & 100 & 10  & $ 3.48 \times 10^{-18}$& $ 3.89 \times 10^{-11}$ & $10^{-9}$  \\ % FRAG\_hiR
S7FB2  & B\&A 1999      & 40 &  50 & 10  & $ 5.31 \times 10^{-23}$& $ 3.89 \times 10^{-11}$ & $10^{-9}$  \\ \hline % FRAG2
S8\_S2   & B\&A 1999      & 15 & 50  &  2  & $ 3.48 \times 10^{-18}$& $ 4.87 \times 10^{-12}$ & $2\times 10^{-10}$  \\ % SIG\_2
S9\_S100 & B\&A 1999      & 27 & 50  &100  & $ 3.48 \times 10^{-18}$& $ 3.11 \times 10^{-10}$ & $10^{-8}$  \\ \hline %SIG\_100
\end{tabular}
\end{center}
\caption[Complete list of all simulations]{Complete list of all simulations (the names of the models are therefore preceded with an ``S''). The first group examines different collisional models, the second group
resumes the nominal simulation S1FB with different resolutions and the third group explores different surface
densities. \label{ComDetail}}
\end{table*}

\begin{enumerate}

\item {\bf Different collision models:} This represents a fundamental
uncertainty, since the impact physics of planetesimals is not well established
yet. In order to do realistic models of planetesimal collisions we need to
understand the internal structure of the bodies taking part in the collision.
Planetesimals emerge as fragile dust aggregates and evolve into solid bodies,
so that their internal structure and strength is time-dependent.

\item {\bf Spatial (radial) density structure} (e.g. gap formation) This is
related to the slowly evolving inhomogeneities introduced by the growing
protoplanets. It has been argued that gap opening in the planetesimal disc
could stop the accretion well before the isolation mass is reached
{\citep[][]{Rafikov2001}}. Our hybrid code includes an accurate treatment of
spatial structuring, so that we are in the position of ascertaining the role of
gap formation in the protoplanetary growth process.

\item {\bf Resolution effects} hinge on the limitation of computing power.
Since the solution of the coagulation equation scales with the third power of
the number of grid cells, the choice of a realistic cut-off mass may be
prohibitively expensive.

\item {\bf Different surface densities:} To address this, we conduct a small
set of different surface densities with our reference fragmentation model
\citep[][impact strength, referred to as B\&A 1999 hereafter]{Benz1999}.

\end{enumerate}

In table~\ref{ComDetail} we summarise the various parameters of our
simulations. In the following subsections we discuss each simulation in more
detail.

We project the {$N-$body} data on to an extended mass grid derived from the
statistical model to generate a unified representation of a hybrid run. This is
so because the hybrid code uses both a statistical representation and $N-$body
data to integrate the planetesimal disc.

\begin{table}
\begin{center}
\begin{tabular}{|r|l|l|l|l|l|l|} \hline
No.       & A & B & C & D & E & F \\ \hline
$T[$yr$]$ & 0 & 1,000 & 10,000 & 20,000 & 50,000 & 100,000  \\ \hline
\end{tabular}
\end{center}
\caption[Time coding]{Integration times from the evolutionary stages A to F \label{TabCode}.}
\end{table}

\subsection{Fragmentation models}

The treatment of collisions is a key element in any simulation of planetesimal
growth. In this section we explore different collisional models with four
different setups, so as to analyse its influence on the final results.

The perfect merger assumption (S3FN hereafter, see table~\ref{ComDetail}) is
the simplest approach for mutual collisions among smaller planetesimals. This
rather simplistic approach can be envisaged as a way to derive an upper limit
for the growth speed in our models.  The second and third model use our
detailed collisional model (see section ``Collisional and fragmentation model''
of Paper I) with the B\&A 1999 impact strength (S1FB) and the approach of
\cite{Housen1990} for the impact strength (S2FH from now onwards).  These two
approaches roughly delimit the range of possible values \citep[see e.g. the
overview in][]{Benz1999}.

The fourth model (S4FBN) assumes that the gaseous disc has dispersed early, so
that we have a gas-free system. This model provides us with a different
evolution for the random velocities, which leads to a different role of the
collisions All other simulations neglect the dispersion of the gaseous disc,
since the simulation time is still short compared to the disc lifetime.

\begin{figure}
\resizebox{\hsize}{!}
          {\includegraphics[scale=1,clip]{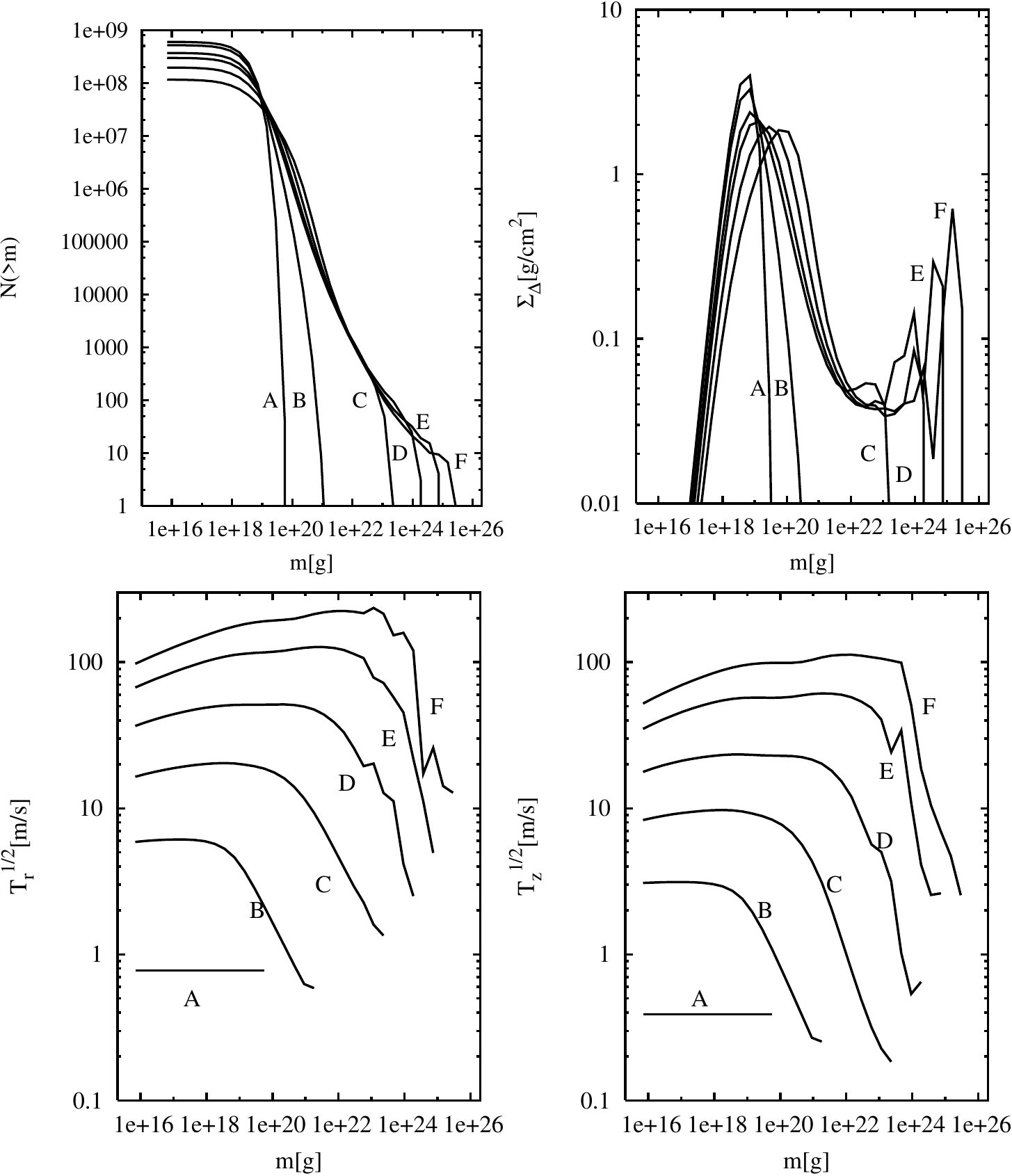}}
\caption
   {
Summary of simulation S3FN, which assumes perfect mergers.
Table~\ref{TabCode} gives the time coding of the labels A--F.
   }
\label{S_NoFrag}
\end{figure}

\begin{figure}
\resizebox{\hsize}{!}
          {
          \includegraphics[scale=1,clip]{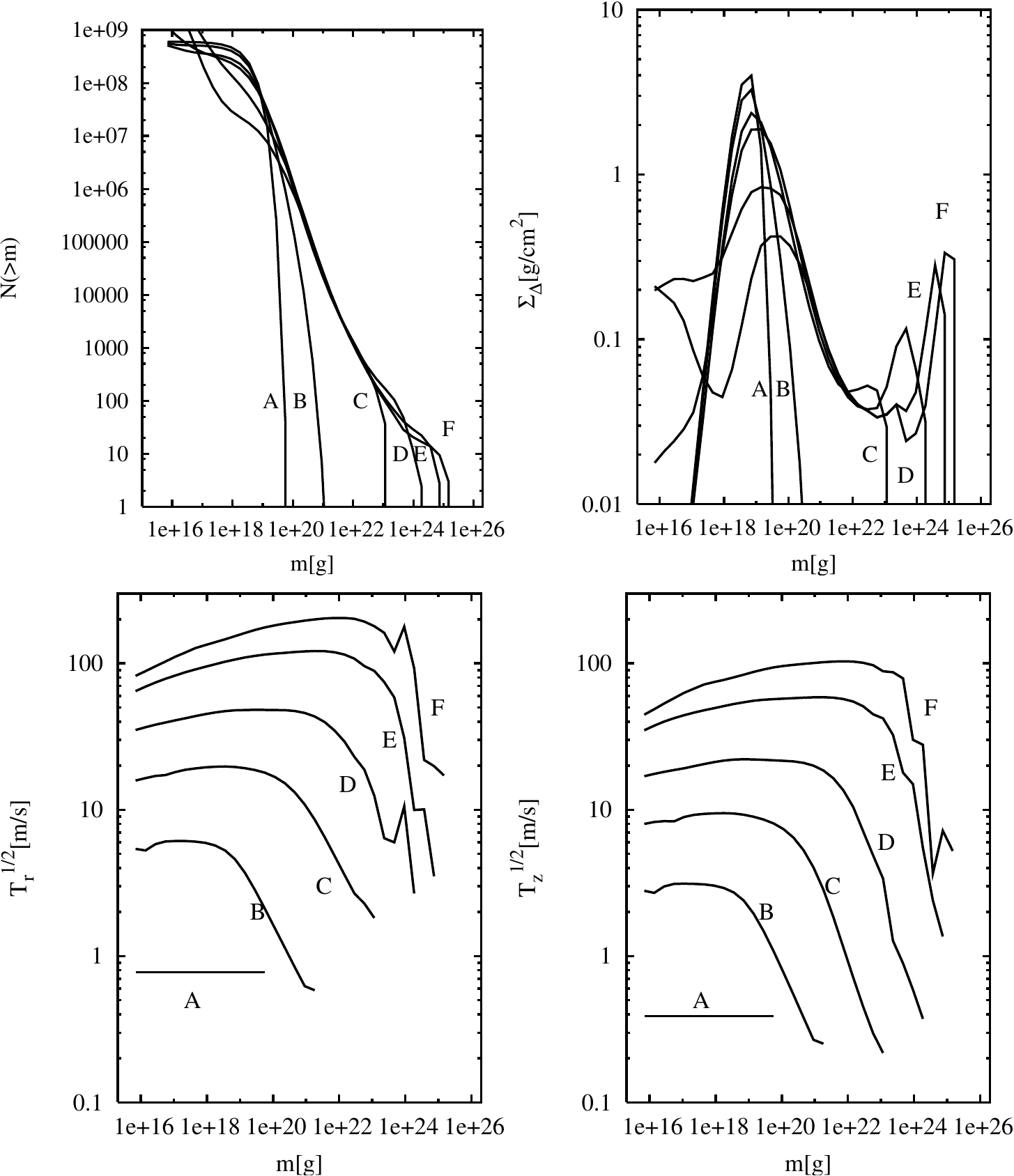}
          }
\caption
   {
Summary of simulation S2FH, which uses the H\&H 1990 strength.
Table~\ref{TabCode} gives the time coding of the labels A--F.
   }
\label{S_Housen}
\end{figure}

\begin{figure}
\resizebox{\hsize}{!}
          {\includegraphics[scale=1,clip]{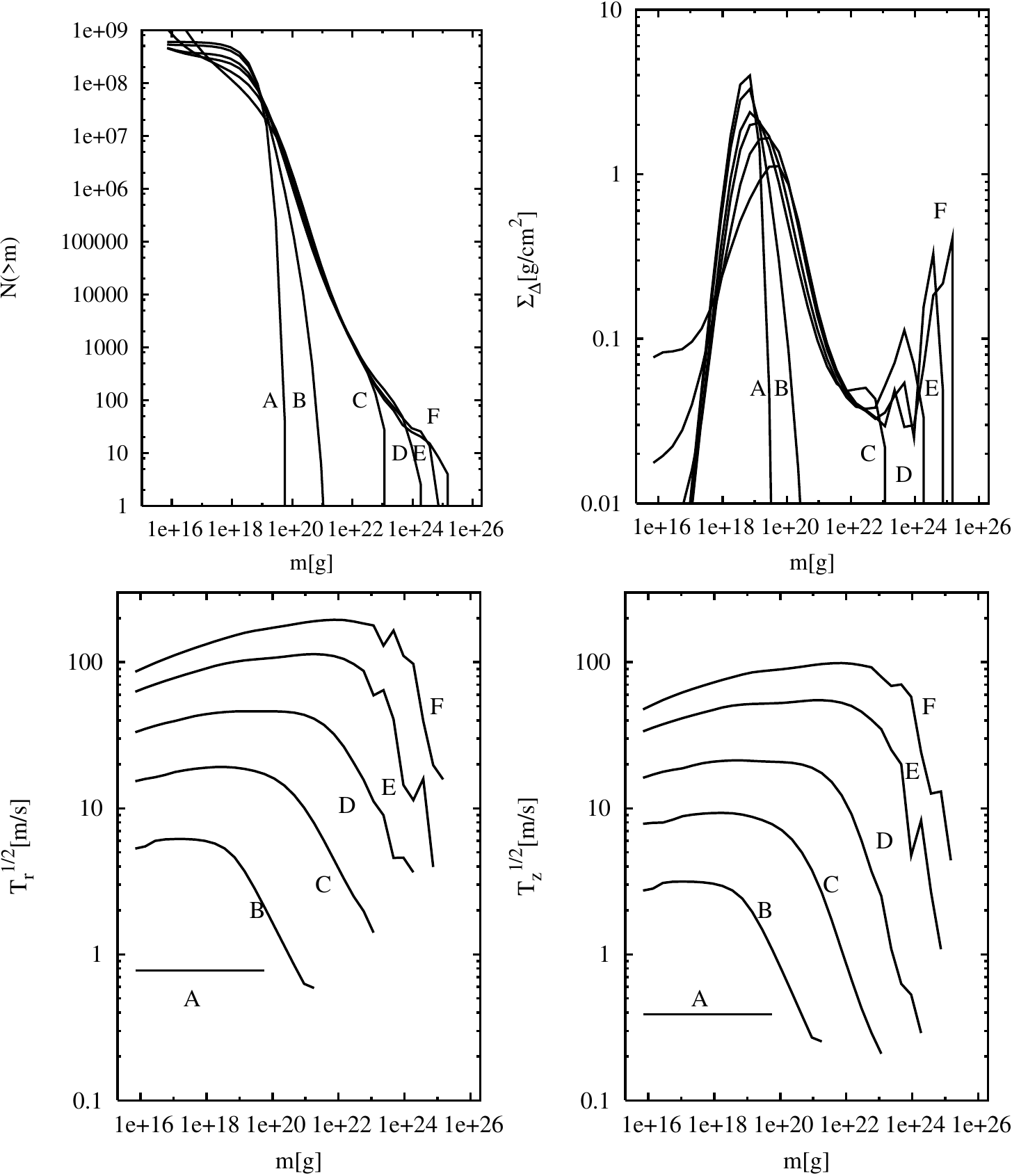}}
\caption
   {
Summary of simulation S1FB, which uses the B\&A 1999 strength.
Table~\ref{TabCode} gives the time coding of the labels A--F.
   }
\label{S_Benz}
\end{figure}

\begin{figure}
\resizebox{\hsize}{!}
          {\includegraphics[scale=1,clip]{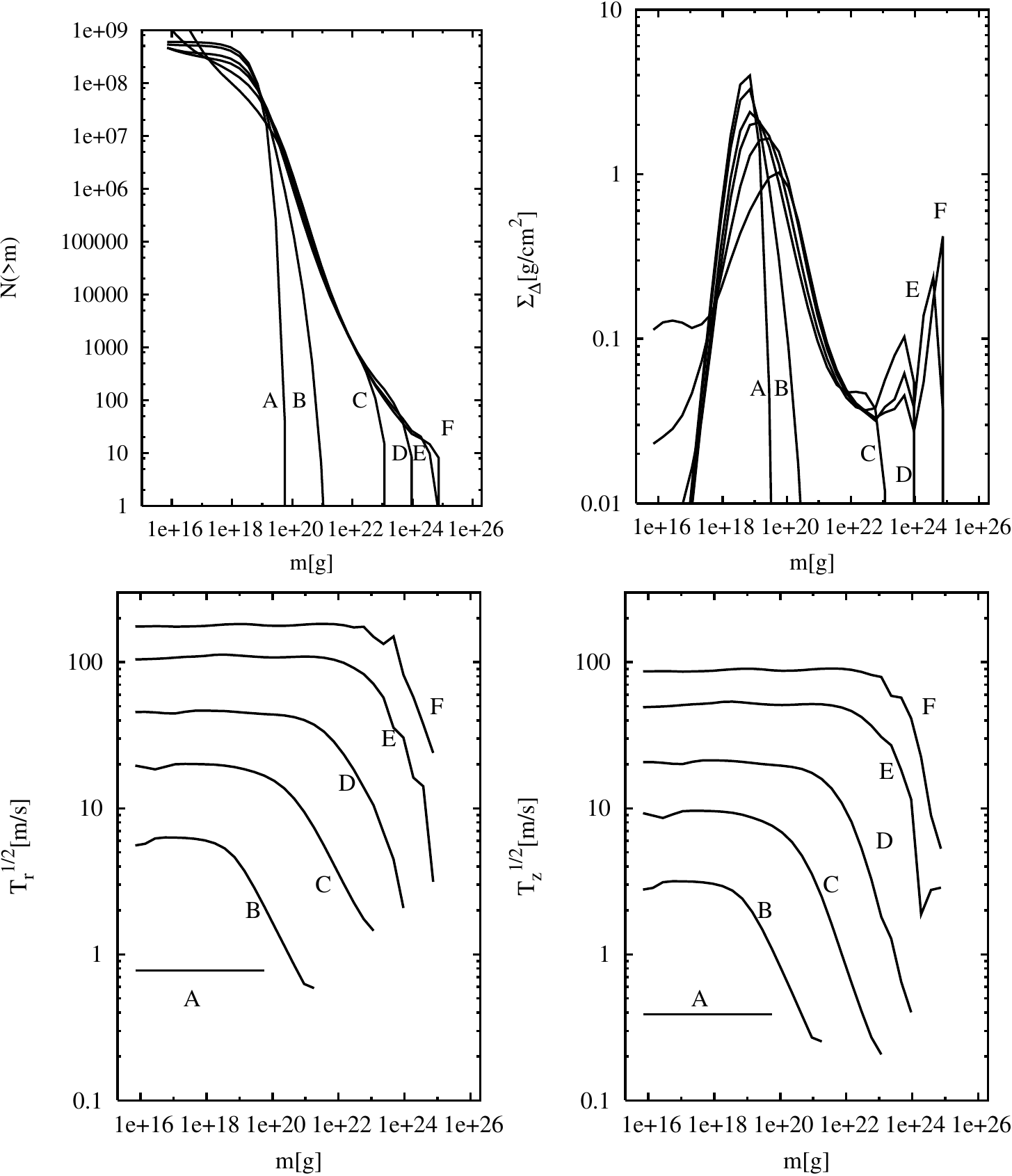}}
\caption
   {
Summary of simulation S4FBN, which uses the B\&A 1999 strength and a gas--free system.
Table~\ref{TabCode} gives the time coding of the labels A--F.
   }
\label{S_NOGAS}
\end{figure}

We present the results of the simulations of the four different approaches in a
figure with four panels: Figure~\ref{S_NoFrag} shows model S3FN,
figure~\ref{S_Housen} model S2FH, figure~\ref{S_Benz} model S1FB and
figure~\ref{S_NOGAS} model S4FBN. In these figures we depict in the upper, left
panel the cumulative size distribution $N(>m)$, which allows us to see the
distribution of particles as a function of the range of masses at different
moments of the integration ($T=0, 10^3, 10^4, 2\times10^4, 5\times10^4 {~\rm
and~} 10^5$ yrs, and we follow in the figures the notation of table~\ref{TabCode}).

On the upper right panel we display the evolution of the surface density per
bin $\Sigma_{\Delta}$. Since we are using a logarithmically equal spacing of
the mass grid, $\Sigma_{\Delta}$ is related to the differential surface density

\begin{equation}
 \Sigma_{\Delta}  \approx  \frac{2}{3}  \frac{\partial \Sigma}{\partial \ln(m)},
\end{equation}

\noindent
where we assume $\delta=2$.

The lower left and right panels show the radial $(T_r)$ and vertical $(T_z)$
velocity dispersion of the system at the different times of
table~\ref{TabCode}.

One conclusion that we can derive immediately in view of these figures is that
in spite of the rather different initial approaches of the models, their time
evolution is rather similar.  The runaway growth sets in after some $10^4$
years, i.e. around stage C in the figures. This is relatively easy to see
because of the pronounced peak at the high mass end. The onset of runaway
growth roughly coincides with the creation of the first $N-$body particles.
Contrary to previous work done with statistical calculations
\citep{Wetherill1989,Wetherill1993}, we find in our models no gap in the size
distribution, but a smooth transition from the slowly growing field
planetesimals (peak around $10^{19}$ g) to the rapidly growing protoplanets.

The initiation of runaway growth is associated with a qualitative change in the
velocity dispersion. While the initial choice of the velocity dispersion
quickly relaxes to a constant value at smaller sizes (transition stage
A$\rightarrow$B), dynamical friction establishes energy equipartition among the
larger masses \citep[see e.g.][in the context of stellar dynamics]{KhalEtAl07}.
The turnover point between these two regimes refers to a balance between the
stirring due to larger bodies and damping due to encounters with smaller
planetesimals \citep[]{Rafikov2003d}.  In addition, the smaller planetesimals
are subjected to damping by the gaseous disc, which significantly reduces the
velocity dispersion at smaller sizes.  Hence this damping is absent in the
gas-free case, which can be seen by comparing the flat distribution of S4FBN,
figure~\ref{S_NOGAS} bottom, with the other models.

We emphasise that all simulations do not generate any artifacts which could be
attributed to an improper joining of the statistical and the $N-$body
component. Some non-smooth structure is visible at the high mass end (i.e., it
is related to data from the $N-$body component), but these variations do
not exceed the fluctuations that we can expect from small number statistics.

All simulations with destructive collisions exhibit the evolution of a
fragment tail. The expected equilibrium slope is roughly $k\approx 2$ (see
section ``Collisional cascades'' of Paper I), which refers to a steep size
distribution and a rather flat density distribution:

\begin{align}
N(>m) &\propto   m^{-1} \nonumber \\
 \Sigma_{\Delta} & \approx   {\rm constant} \label{AprEqSig}
\end{align}

\noindent
Simulation S1FB (B\&A 1999 strength, figure~\ref{S_Benz}) and S4FBN (gas--free,
figure~\ref{S_NOGAS}) show a clear plateau in the density distribution around 0.1, in
accordance with the previous estimate, Eq.~\ref{AprEqSig}.  In contrast,
simulation S2FH (H\&H 1990 strength, figure~\ref{S_Housen}) evolves a second
maximum at the lower boundary of the mass grid. Although this structure is partly
due to the lower grid boundary, the main cause is the reduced H\&H 1990 impact
strength at sizes of a few 10 kilometres (as compared to the B\&A 1999
strength), which leads to the quick destruction of the remaining field
planetesimals at masses around $10^{18}$ g.

\begin{figure}
\resizebox{\hsize}{!}
          {\includegraphics[scale=1,clip]{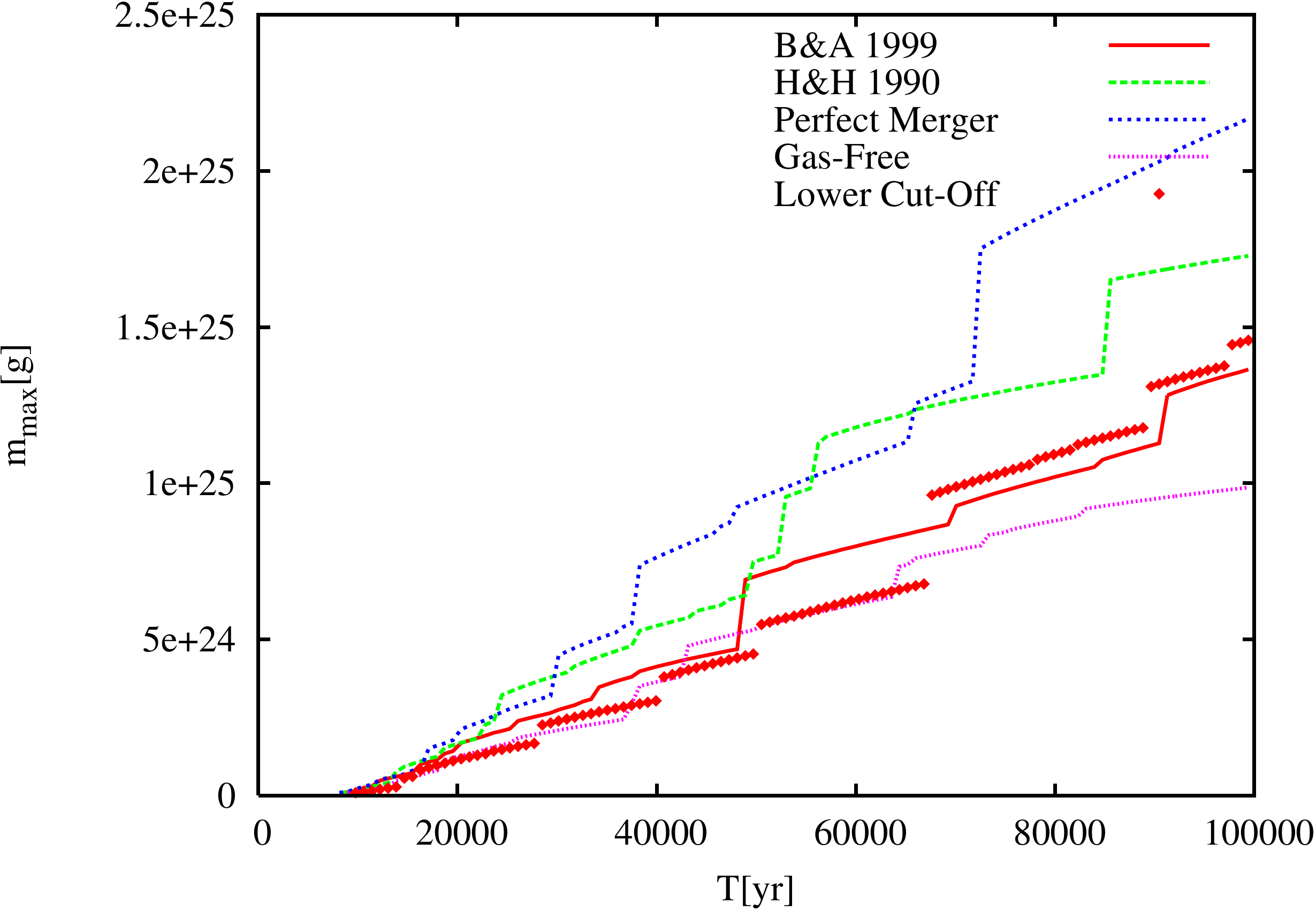}}
\caption
   {
 Largest body in the simulation as a function of
time for the different collision {models} S1FB (B\&A 1999), S2FH (H\&H 1990), S3FN (perfect merger) and S4FBN (gas-free).
In {addition}, we also include simulation S7FB2 with a lower cut-off mass.
   }
\label{Mmax_Coll}
\end{figure}

\begin{figure}
\resizebox{\hsize}{!}
          {\includegraphics[scale=1,clip]{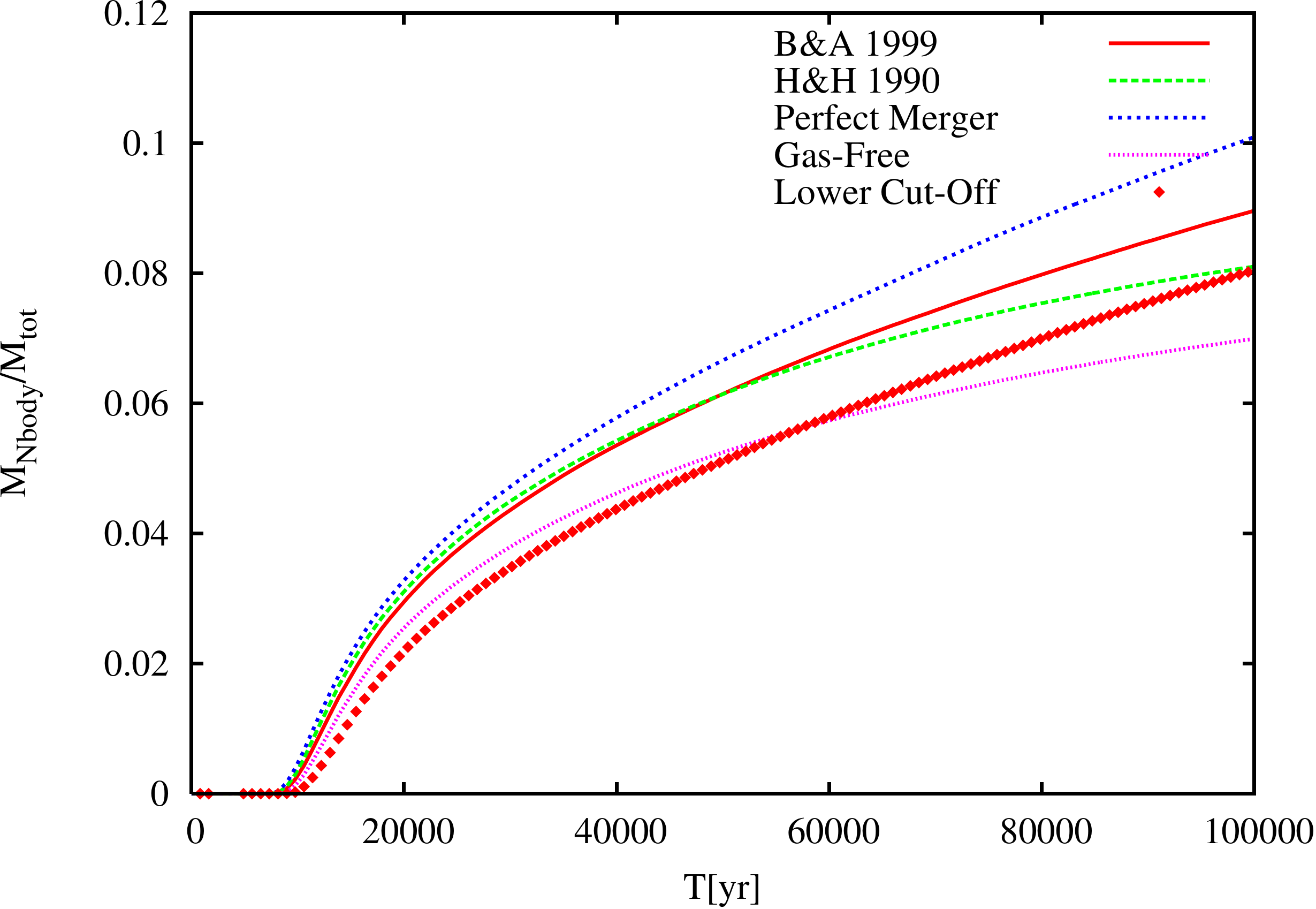}}
\caption
   {
 The same as figure~\ref{Mmax_Coll} for the total mass in the $N-$body component.
   }
\label{Mnb_Coll}
\end{figure}

The overall agreement of the different simulations is reflected by the growth
of the largest mass in the system, as we depict in figure~\ref{Mmax_Coll}.  Up
to $2\times 10^4$ years, all simulations agree well. Later on simulation S3FN
(which follows the approximation of perfect mergers) exhibits the largest
growth rate, as one would naturally expect. Although simulation S1FB (which
uses the B\&A 1999 strength prescription) seems to show a slower growth than
simulation S2FH (which follows the recipe of H\&H 1990 for strength), this is
only due to a different sequence of major impacts. In fact, the B\&A 1999
strength simulation makes possible a much faster growth, in accordance with the
total mass contained in the $N-$body component, which is displayed in
figure~\ref{Mnb_Coll}. The gas-free simulation S4FBN exhibits the slowest
growth among the four test cases.

A further examination of the mass loss -- which we define as the mass in
planetesimals which cross the lower grid boundary-- reveals the cause of this
different behaviour: A pronounced mass loss in simulation S2FH slows down the
protoplanetary growth reducing the surface density.  In the gas-free
case, the accretion rate is mainly reduced because of a larger velocity
dispersion, although we can still notice some enhanced mass loss by comparing the
lower panels of figures~\ref{S_Benz} and \ref{S_NOGAS}.

We find no accelerated growth due to the inclusion of fragmentation events,
contrary to the work of \cite{Wetherill1989}. We find that a lower impact strength or the
absence of gas damping slows down the growth by an increased mass loss.  The
total mass in the $N-$body component is still small at the end of the
simulations, of about $\approx 10\%$ of the total mass, as shown in
table~\ref{PropMax} of next section.

\subsection{Spatial distribution \label{SpatDist}}

\begin{figure*}
\resizebox{\hsize}{!}
          {\includegraphics[scale=1,clip]{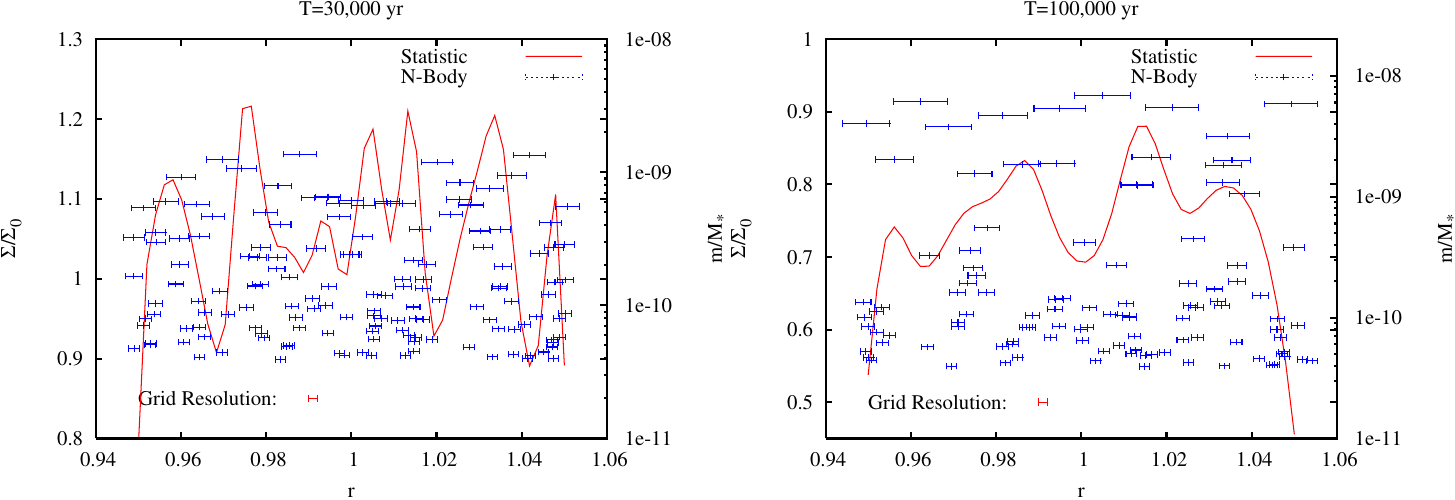}}
\caption
   {
{\em Left y-axis and solid, red curve of the left panel:} Radial density structure of the
statistical component of model S1FB at $T=3\times10^4$~yr.  {\em Right y-axis
and blue dots of the left panel:} Semimajor axis and masses of the $N-$body particles in the
simulation after the same amount of time.  The error bars are 10 Hill radii
wide and refer to the heating zone of each $N-$body particle. We also display the
grid resolution as a reference point.
In the right panel we depict the same after $T=10^5$ yrs.
   }
\label{Gap1_30_100}
\end{figure*}

\begin{figure}
\resizebox{\hsize}{!}
          {\includegraphics[scale=1,clip]{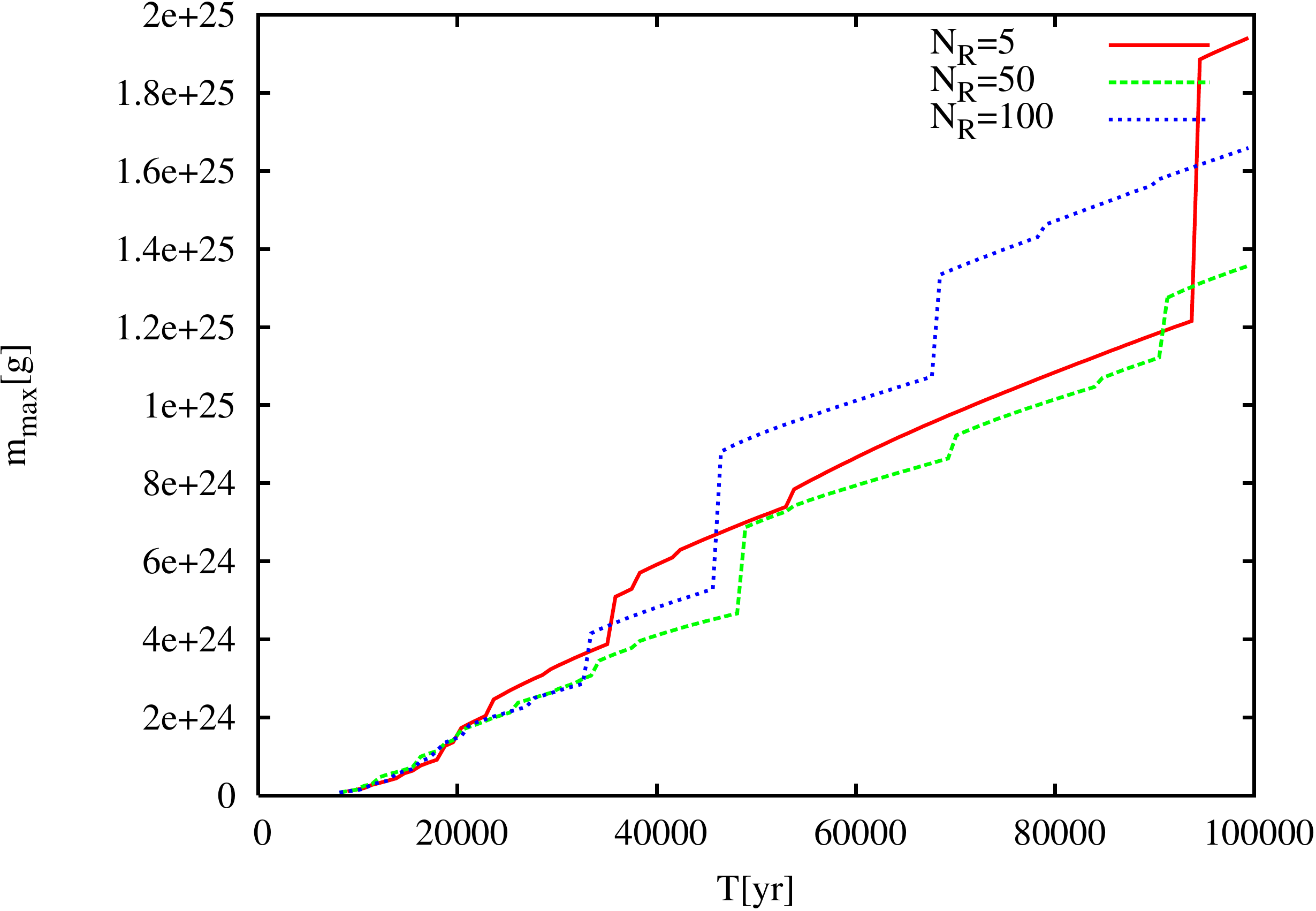}}
\caption
   {
Largest body in the simulation as a function of
time for the different resolutions S5FBL $(N_R=5)$, S1FB $(N_R=50)$ and S6FBH $(N_R=100)$.
   }
\label{Mmax_Res}
\end{figure}

How well the code can treat spatial inhomogeneities depends on the choice of
the spatial resolution. We hence compare a low resolution model, model S5FBL of
table~\ref{ComDetail}, which virtually inhibits any spatial structuring, with a
model that uses our fiducial resolution, model S1FB, as well as with a model
that has a finer resolution, S6FBH. We adjust the fiducial resolution to the
width of the heating zone of a planetesimal at the transition mass.

In the left panel of figure~\ref{Gap1_30_100} we have the spatial structure at
$T=30,000$ yr, i.\,e. shortly after stage D (nominal model S1FB). While the
protoplanets are already massive enough after a few $10^4$ years (stage C) to
open gaps in the planetesimal component, there is only a weak correlation
between the radial structures and the location of the most massive
protoplanets. A closer examination of the time evolution of the radial
structure reveals that most features are ``fossils'' from the first emerged
$N-$body particles, which are slowly washed out by the diffusion of the field
planetesimals. In the right panel of the same figure~\ref{Gap1_30_100} we
confirm the further smoothing of the radial features. While major mergers among
the protoplanets still lead to distinct features in the surface density even
after a few $10^4$ years, any further structuring ceases at the end of the
simulation.

The absence of any prominent gap formation (fluctuations are smaller than 20
\%) is related to the evolution of the overall size distribution. Though the
gap opening criterion (see section ``Protoplanet growth'' of Paper I, and we
reproduce here the relevant equation for convenience)

\begin{eqnarray}
 \frac{M_{\mathrm{gap}}}{M_c} & \approx &
 \left\{
 \begin{array}{lll}
 \frac{\Sigma a^2}{M_c} \left(\frac{m}{M_c}\right)^{{1}/{3}} & \mbox{if} & v \lesssim \Omega r_{\mathrm{Hill}} \\
 \frac{\Sigma a^2}{M_c} \left(\frac{m}{M_c}\right)^{1/3}\left(\frac{\Omega r_{\mathrm{Hill}}}{v}\right)^2 & \mbox{if} & v \gg \Omega r_{\mathrm{Hill}},
 \end{array}     \right. \label{Mgap}
\end{eqnarray}

\noindent
is formally satisfied by all protoplanets during the runaway phase, the dense
overlapping of the associated heating zones (see figure~\ref{Gap1_30_100})
inhibits the evolution of any gap-like feature. As the protoplanets grow, they
exert a growing influence on the dynamics of the planetesimal system. While
this dominance could in principle enhance gap formation, the system is
already dynamically too hot to allow the system to develop radial structures. The eccentricities of
the field planetesimals are comparable to the width of the heating zone
(compare figure~\ref{S_Benz}, bottom), and hence any planetesimal that is scattered
to larger (or smaller) radii immediately encounters a neighbouring protoplanet.

In summary, the protoplanets (or rather their precursors) are too abundant when
the system is dynamically cool enough, but when a group of mature protoplanets
has evolved, the system is already too hot. Thus, we expect an even less
effective radial structuring for larger surface densities. While systems with a
lower surface density may lead to the formation of gap-like structures, they
evolve slowly that planet formation may never reach the final growth
phases.

\subsection{Resolution}

We can further evaluate the (minor) role of gap formation by comparing the
growth process for the three different radial resolutions  $N_R=5$, $N_R=50$
and $N_R=100$. Besides some variations due to a different sequence of major
impacts (see figure~\ref{Mmax_Res}), all three simulations are in excellent
agreement with respect to the mass loss and the total mass in the $N-$body
component.

Accordingly, we find no differences between the various fragmentation models
(S1FB, S2FH, S3FN, S4FBN) with respect to possible emerging gaps, except an
earlier homogenisation in the gas-free case S4FBN due to the stronger heating
of the smaller planetesimals.

\begin{figure}
\resizebox{\hsize}{!}
          {\includegraphics[scale=1,clip]{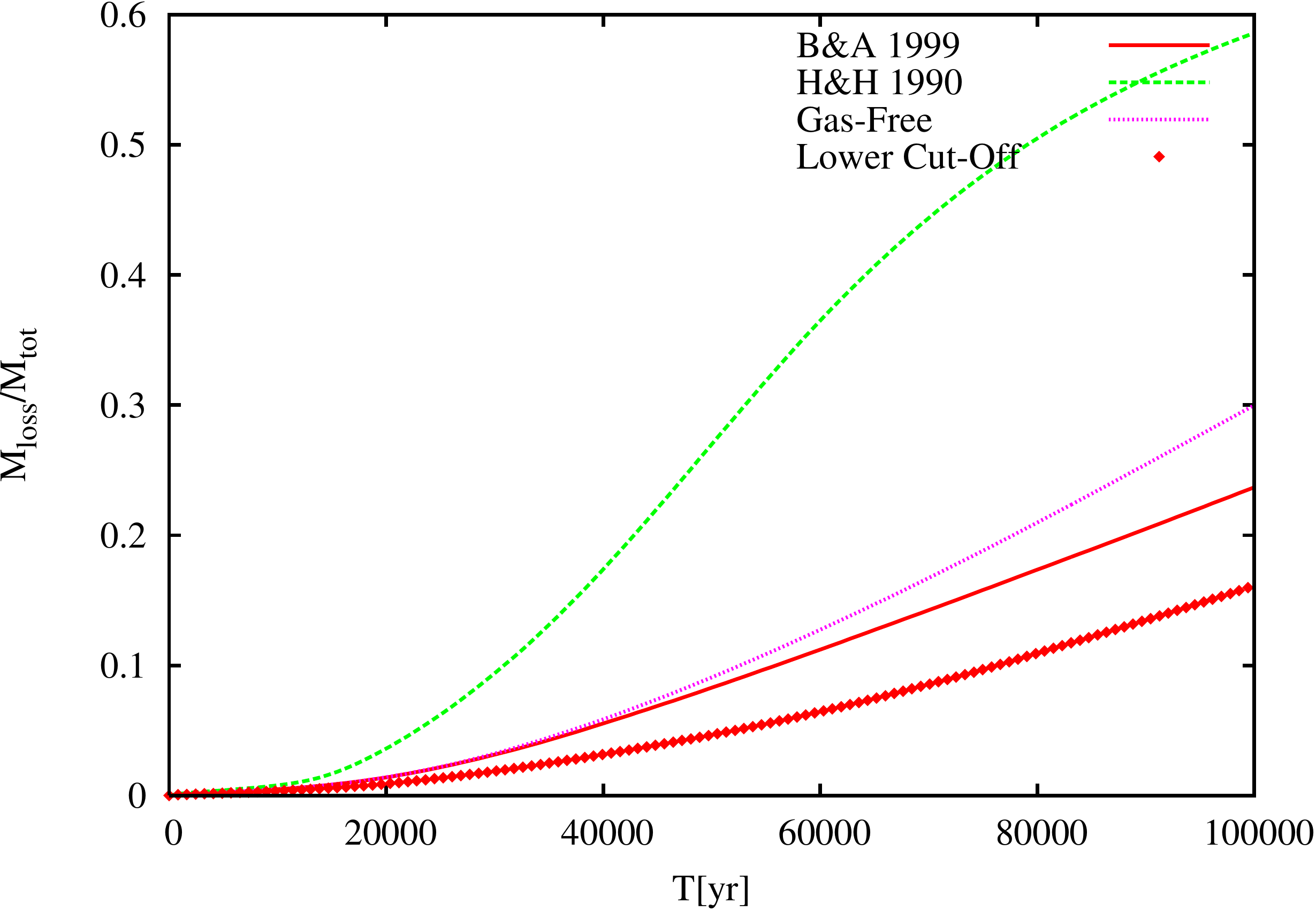}}
\caption
   {
 Same as in figure~\ref{Mmax_Coll} for the total mass loss. For obvious reasons we
do not include S3FN, which assumes perfect merger.
   }
\label{Mloss_Coll}
\end{figure}

\begin{figure}
\resizebox{\hsize}{!}
          {\includegraphics[scale=1,clip]{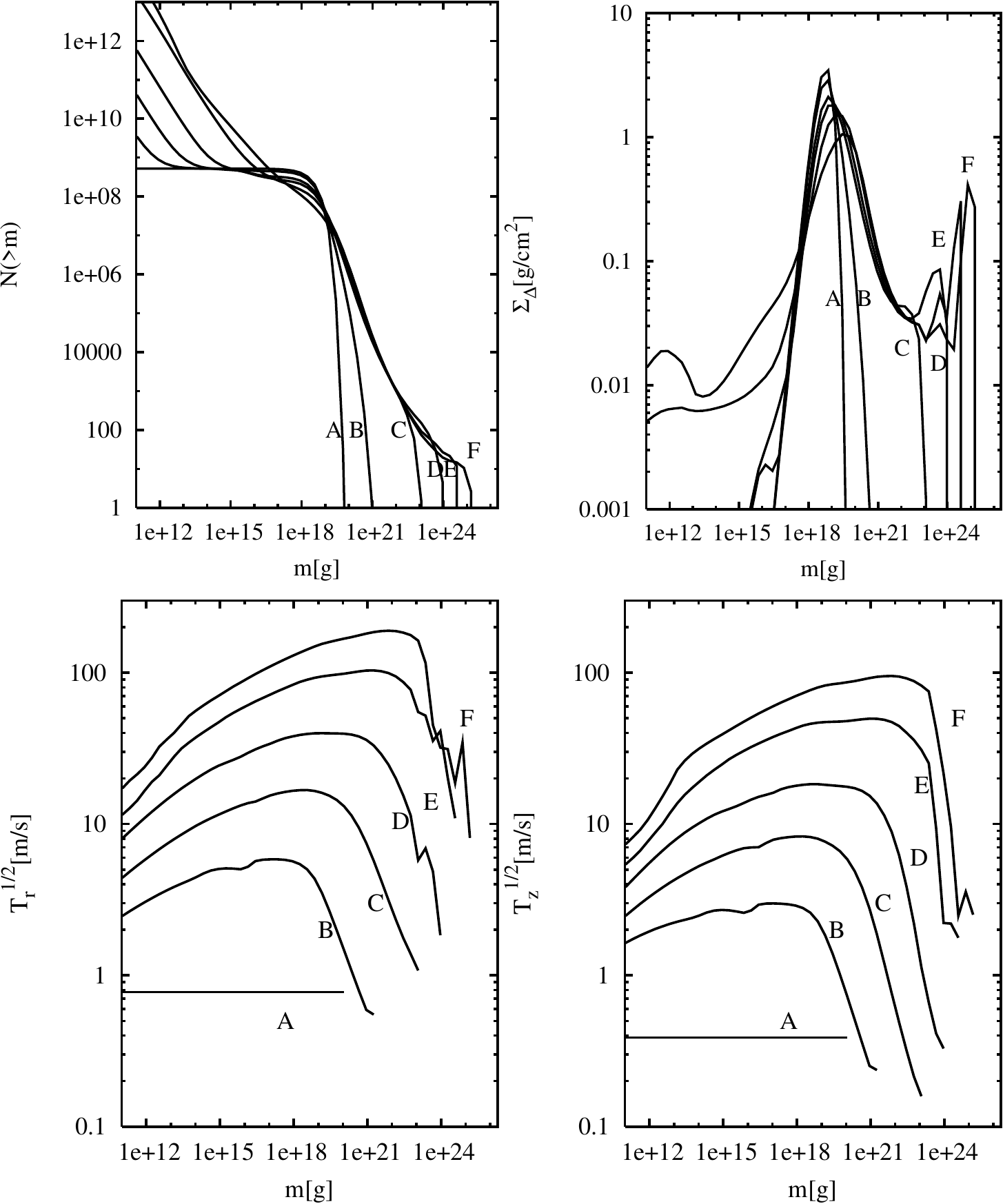}}
\caption
   {
Summary of simulation S7FB2, which uses the B\&A 1999 strength and a smaller
lower cut-off mass.
   }
\label{S_Benz2}
\end{figure}

We conduct one additional simulation, named S7FB2, see figure~\ref{S_Benz2}, in
which we reduce the lower mass grid boundary by a factor $10^5$.  Although the
standard choice $m_{\min}=6.9 \times 10^{15}$ g is in accordance with the size
regime where migration would remove the smaller fragments, the actual mass
cut-off is less sharp as we estimated in section ``Collisional cascades'' of
Paper I.  A reduced lower cut-off increases the dwell time of collisional
fragments in the system, thus increasing the mass fraction which could be
accreted by the protoplanets. As a result, mass loss is reduced by 30\%
compared to our fiducial case S1FB, as we can see in figure~\ref{Mloss_Coll}.
AltThough the shape of the fragment tail is modified by a different choice of
the grid boundary, the change of the overall evolution of the protoplanets is
rather small.

\subsection{Surface density}

\begin{table*}
\begin{center}
\begin{tabular}{|l|c|c|c|c|c|} \hline
Simulation & $\Sigma[$g$/$cm$^2]$ & m$[$g$]$ & $v_{\mathrm{Hill}}[$m$/$s$]$ & $M_{\mathrm{Nbody}}/M_{\mathrm{Statistic}}$ & $M_{\mathrm{iso}}$[g] \\ \hline
S8\_S2   &   2 & $2.6 \times 10^{23} $ &   10.5  & 0.04 & $7.8\times 10^{25}$   \\ \hline
S1FB     &  10 & $1.2 \times 10^{25} $ &   37.6  & 0.13 & $8.6\times 10^{26}$   \\ \hline
S9\_S100 & 100 & $4.1 \times 10^{26} $ &  122.2  & 1.34 & $2.7\times 10^{28}$   \\ \hline
\end{tabular}
\end{center}
\caption[Maximum mass and associated quantities]
{Maximum mass and associated quantities at $T=100,000$ years for different surface densities.\label{PropMax}}
\end{table*}

We now examine the evolution of different surface densities with a last set of
simulations.  We take S1FB as our nominal model, with a a surface density of
$\Sigma = 10$ g$/$cm$^2$.  We explore two different surface densities: A
low--mass disc with $\Sigma = 2$ g$/$cm$^2$ (simulation S8\_S2), and a
high--mass disc with $\Sigma = 100$ g$/$cm$^2$ (simulation S9\_S100), which is
close to the upper mass limit set by observations.  The basic parameters of all
three simulations are equal except a proper scaling of the gas density and
transition masses chosen individually.

We first resume the inspection of possibly emerging gaps: While the low
mass case
shows a more pronounced radial structure (with fluctuations as large as 40\%),
these features are only weakly related to the location of the largest protoplanets.
Hence, these structures are signatures of the first emerging {$N-$body} particles.
The high mass case exhibits no strong features at all, except for very weak features
during the initial runaway phase.
These results strengthen the discussion in Section~\ref{SpatDist}, assigning
only a minor role to gap formation in the planetesimal component during the protoplanet
accretion.

\begin{figure}
\resizebox{\hsize}{!}
          {\includegraphics[scale=1,clip]{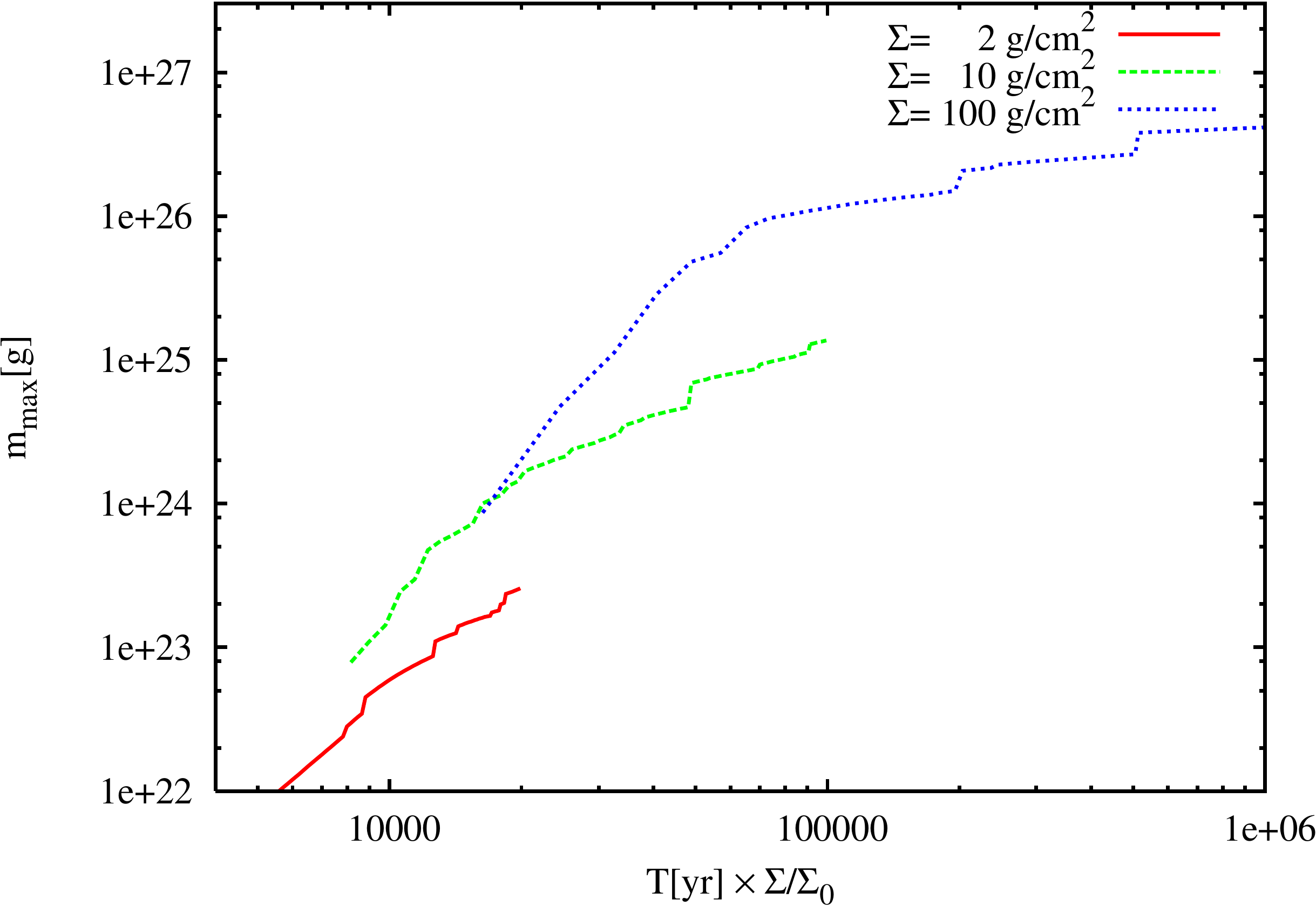}}
\caption
   {
Largest body in the simulation as a function of
time for the different surface densities S8\_S2 ($\Sigma = 2$ g$/$cm$^2$), S1FB ($\Sigma = 10$ g$/$cm$^2$)
and S9\_S100 ($\Sigma = 100$ g$/$cm$^2$). The reference density is $\Sigma_0 = 10$ g$/$cm$^2$.
   }
\label{Mmax_SIG}
\end{figure}

\begin{figure}
\resizebox{\hsize}{!}
          {\includegraphics[scale=1,clip]{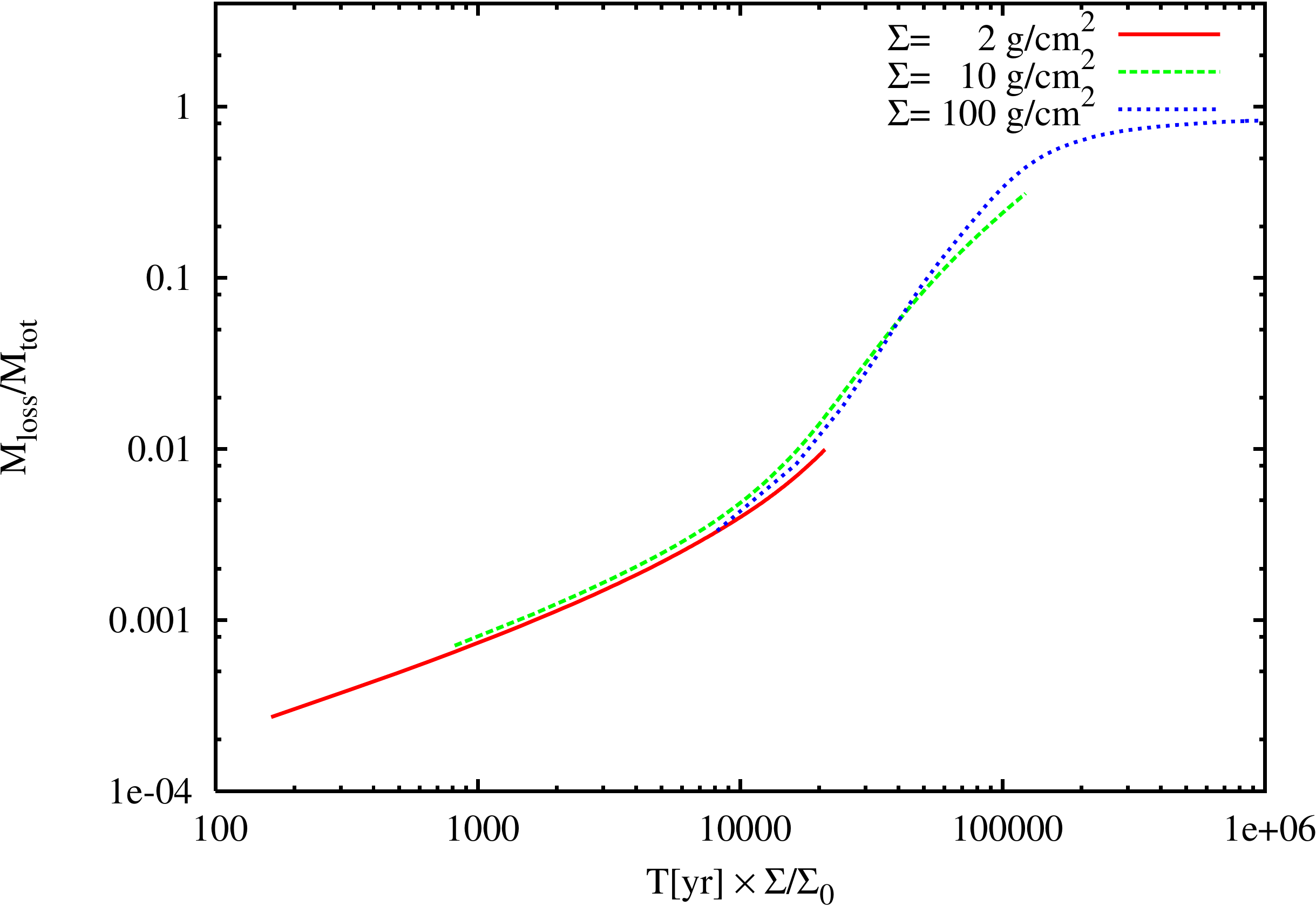}}
\caption
   {
Mass loss in the simulation as a function of
time for different surface densities for the same
cases of figure~\ref{Mmax_SIG}.
   }
\label{Mloss_SIG}
\end{figure}

The overall growth process follows a standard pattern. Since the accretion rate
in all three simulations is directly proportional to the surface density (see
the following equation of section ``Protoplanet growth'' of Paper I), we
rescale the time to the reference simulation S1FB. We obtain a good agreement
in the time evolution of the largest mass in the system, as we can see in
figure~\ref{Mmax_SIG}, although the turnover to the slower oligarchic growth
occurs at different (scaled) times. Likewise, we rescale the time to ease the
comparison of the mass loss in the three simulations, which we depict in
figure~\ref{Mloss_SIG}.

As soon as a set of dominant protoplanets has evolved, they control the
velocity dispersion of the field planetesimals. Therefore the magnitude of the
velocity dispersion matches the Hill velocity of the largest body in the system
(see Table~\ref{PropMax} and figure~\ref{S_Benz}, \ref{S_SIG2} and
\ref{S_SIG100}).

While this similarity in the three simulations is also in good agreement with
standard estimations of the growth process (see the section ``Initial models''
of Paper I), the later stages in the evolution differ markedly: A larger
surface density implies larger (and faster growing) protoplanets, so that the
velocity dispersion of the field planetesimals is also driven to higher
velocities.  This therefore leads to an increased mass loss as the initial
surface density increases (figure~\ref{Mloss_SIG}).  The mass loss of the most
massive setup S9\_S100 reduces the surface density nearly to the standard case
S1FB.  Since the mass loss is not due to actual migration of smaller fragments,
but to the lower grid boundary (in mass) which mimics the effect of migration,
this effect deserves a closer examination:

The influence of fragmentation on the protoplanetary growth is mainly
determined by two timescales: The fragmentation time,  $\tau_{\mathrm{frag}}$,
which refers to collisions between planetesimals, and the growth timescale
$\tau_{\mathrm{grow}}$ of the protoplanetary accretion.

\begin{figure}
\resizebox{\hsize}{!}
          {\includegraphics[scale=1,clip]{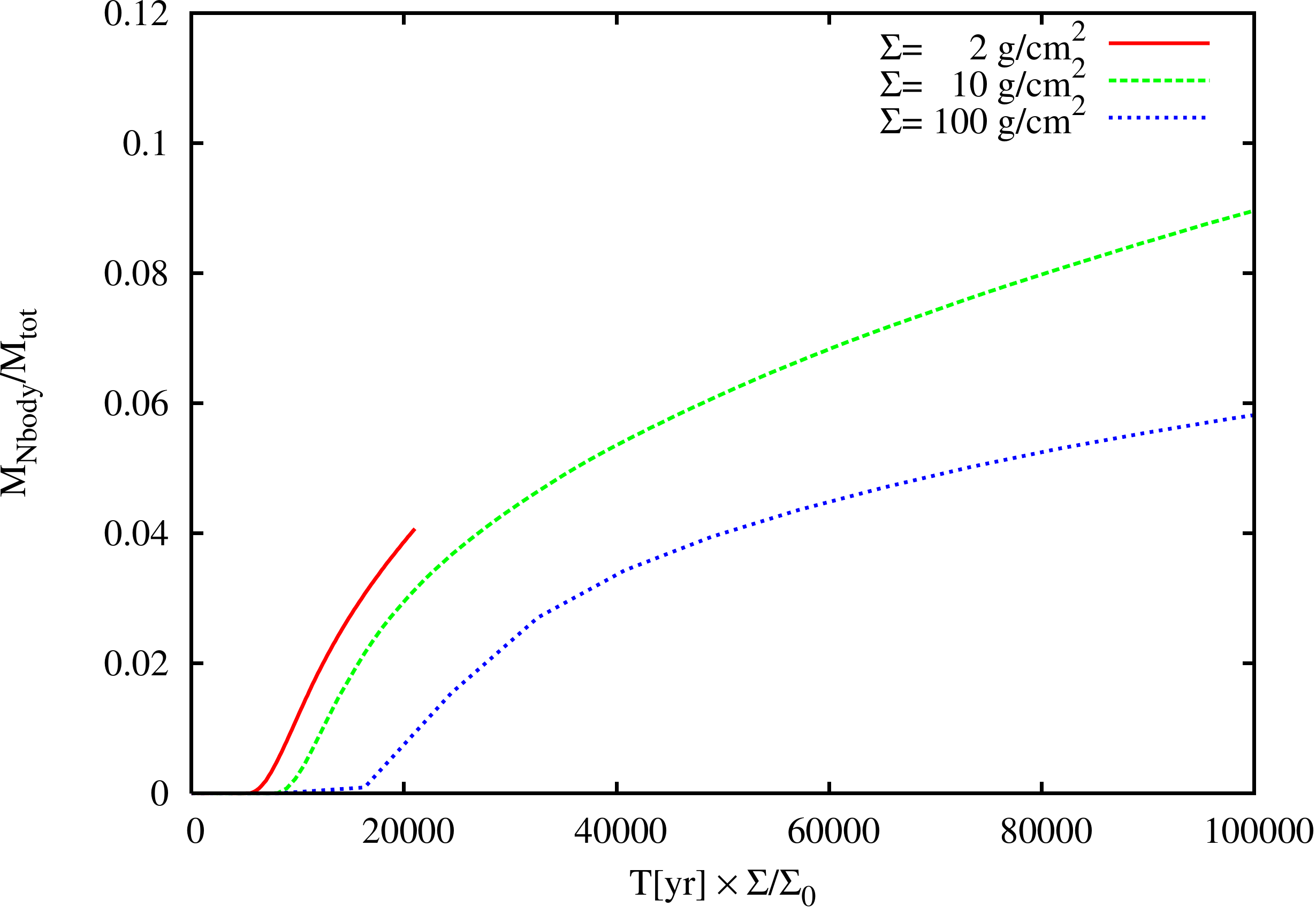}}
\caption
   {
Mass in the $N-$body component of a simulation as a function of
time for the different surface densities S8\_S2 ($\Sigma = 2$ g$/$cm$^2$), S1FB ($\Sigma = 10$ g$/$cm$^2$)
and S9\_S100 ($\Sigma = 100$ g$/$cm$^2$).
   }
\label{Mnb_SIG}
\end{figure}

We employ the expressions derived in section ``Perturbation of equilibrium'' of
Paper I and the approximated differential surface density of equation~\ref{AprEqSig} to
estimate the {fragmentation} time:

\begin{align}
 \tau_{\mathrm{frag}} &\approx  \frac{\tilde S}{G_0'} \tau_0 \qquad G_0' \approx 10 \nonumber \\
   &\approx  \frac{\ln(m/m_0)}{80} \frac{R_m S_m}{\Sigma_m \Omega^3  R_{\mathrm{Hill}}^2 }.
\end{align}

\indent
In the last equation
$m$ is a typical mass of the largest planetesimals, $R_m$ is the corresponding radius and
$S_m$ is the impact strength. $\Sigma_m$ is the total surface density of the field planetesimals,
with a lower cut-off $m_0$ due to migration. $R_{\mathrm{Hill}}$ is the typical Hill radius of
a protoplanet, where it is assumed that the protoplanets control the velocity dispersion of
the field planetesimals.

\section{The mill condition}
\label{sec.mill}

The growth timescale of the protoplanets follows immediately from rearranging the following
equation, as discussed in section ``Protoplanet growth'' of Paper I:

\begin{equation}
\dot M \approx   6\pi\Sigma \Omega \frac{RR_{\mathrm{Hill}}}{\tilde e_m^2} \label{MdotOli}
\end{equation}

\noindent
The timescale hence is

\begin{equation}
 \tau_{\mathrm{grow}}  \approx  \frac{M \tilde e_m^2}{6\pi\Sigma_m \Omega RR_{\mathrm{Hill}} }
\end{equation}

\noindent
Since the mass loss due to migration and the replenishment of smaller fragments
by mutual collisions quickly establishes a stationary solution, the removal of
the field planetesimals operates on the fragmentation timescale.  Since the
protoplanets grind the surrounding planetesimals without retaining a
significant fraction, the accretion of the protoplanet ceases if the condition

\begin{equation}
 \tau_{\mathrm{grow}}  >  \tau_{\mathrm{frag}} \label{MillCond}
\end{equation}

\noindent
is fulfilled, which we will refer to from now onwards as the ``mill condition''.
We can easily derive a lower limit for the protoplanet mass assuming $\tilde e_m =4$
and by translating this condition in terms of mass, the ``mill mass'',

\begin{align}
 M &>  \frac{1}{53} \ln(m/m_0) \frac{R_m S_m}{\Omega^2} \sqrt[3]{\frac{M_c}{a^3\rho}},
\end{align}

\noindent
which we denote as $M_{\mathrm{mill}}$:

\begin{equation}
 \frac{M_{\mathrm{mill}}}{m}  =  \frac{f}{53}\ln(m/m_0) \left( \frac{2S_m}{\rho v_{\infty,m}^2} \right) \left( \frac{a^3 \rho}{M_c} \right)^{2/3} \label{Mmill}
\end{equation}

\noindent

In the last expression $\rho$ is the bulk density of the planetesimals and
$v_{\infty,m}$ is the escape velocity of the field planetesimals.  $f$ is a
factor of order unity to take into account alternative treatments of migration
that could alter the size of $M_{\mathrm{mill}}$.

We note that a necessary condition for the mill process to operate is the
presence of a gaseous disc.  Since a high surface density is needed for the
protoplanetary growth to reach the mill mass, the growth itself is likely to be
faster than the dispersal of the gaseous disc.

Nevertheless, the concept is also useful in a gas-free system: If the
protoplanets in a given planetary system do not exceed the mill mass, it is
still possible that the planets after the final giant impact phase exceed
$M_{\mathrm{mill}}$. Radiative pressure and Pointing--Robertson drag still
provide an effective removal of dust-sized particles in a gas-free system
\citep[see the discussion in][]{Burns1979}; hence, while the absence of strong
migration of planetesimals prevents any reduction of the planetary accretion
rate, the system enters nevertheless a qualitatively different stage: The
evolution of the left-over planetesimals (i.\,e. the {\em disc clearing}) is
now driven by fragmentation rather than accretion.

The mill mass is that {\em is independent of the surface density of the field
planetesimals and hence represents a universal upper limit of the protoplanet
mass}, given that all other parameters of the planetary system are fixed. The
mill mass increases more steeply with radius ($\propto r^2$) than
the isolation mass for any realistic density profiles (e.g. $M_{\mathrm{iso}
}\propto  r^{3/4}$ for the minimum mass solar nebula).

\begin{figure}
\resizebox{\hsize}{!}
          {\includegraphics[scale=1,clip]{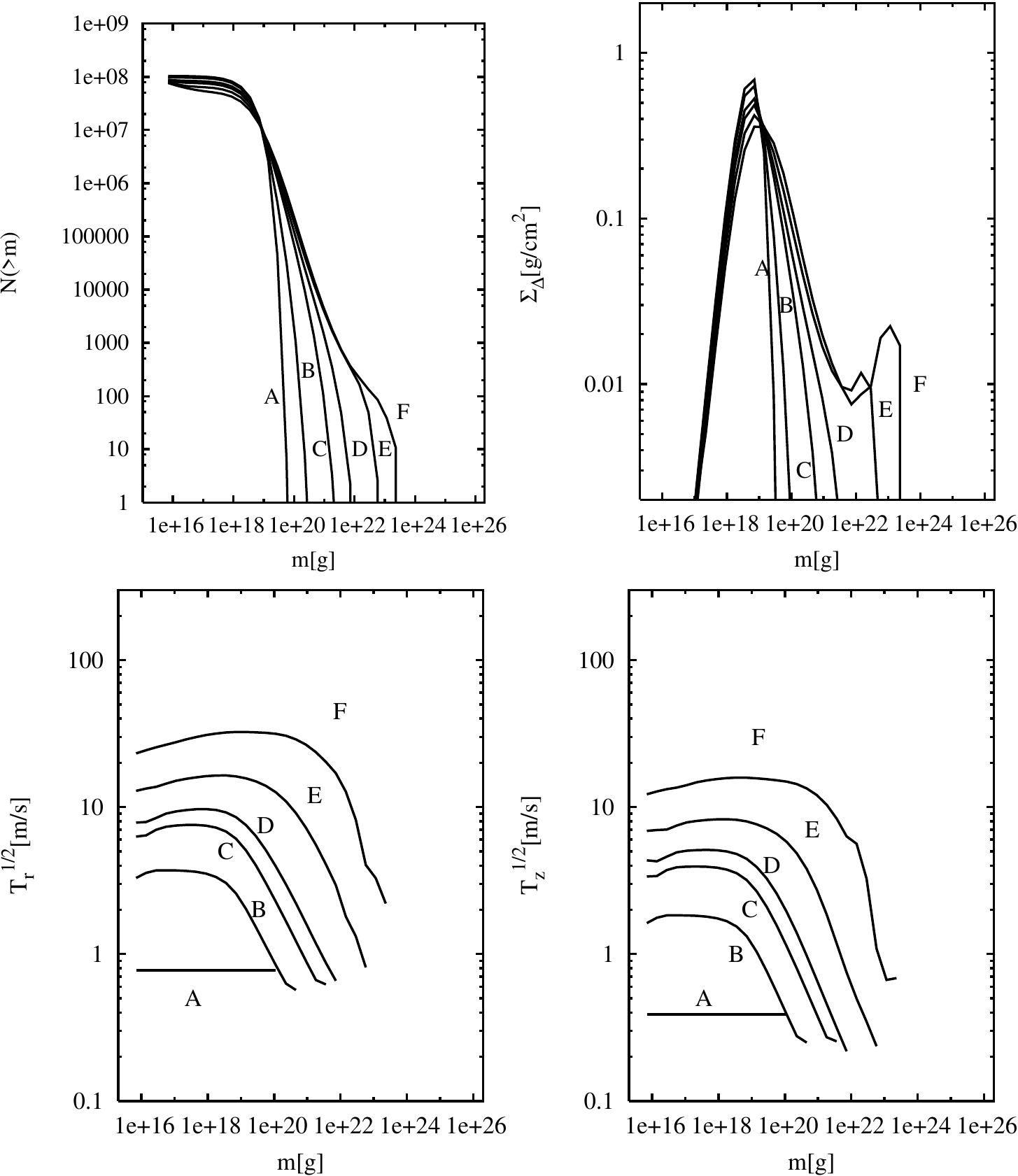}}
\caption
   {
Summary of simulation S8\_S2, which uses the B\&A 1999 strength and a lower
surface density $\Sigma = 2$ g$/$cm$^2$. Table~\ref{TabCode} gives the time coding of the labels A--F.
   }
\label{S_SIG2}
\end{figure}

\begin{figure}
\resizebox{\hsize}{!}
          {\includegraphics[scale=1,clip]{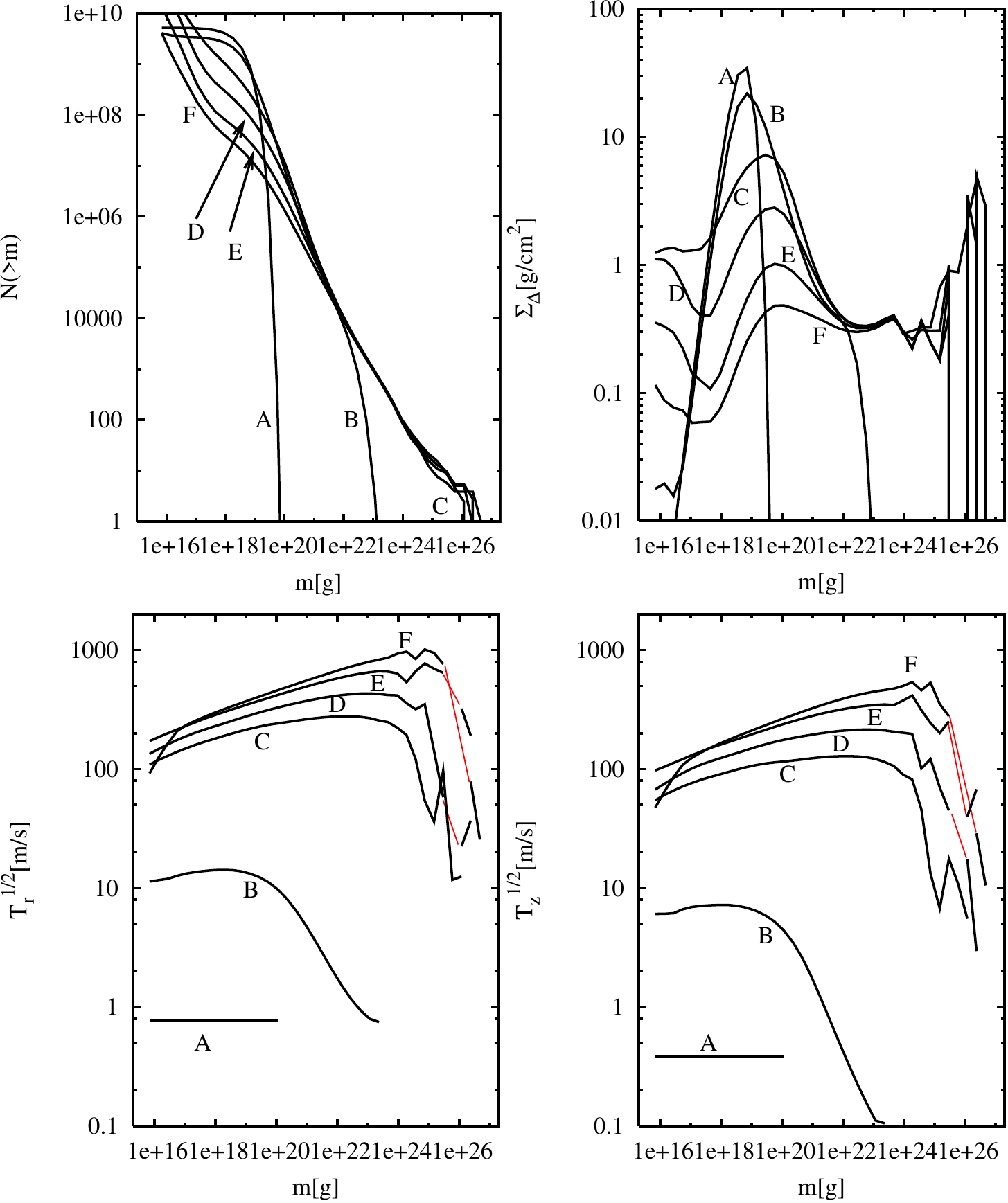}}
\caption
   {
Summary of simulation S9\_S100, which uses the B\&A 1999 strength and a higher
surface density $\Sigma = 100$ g$/$cm$^2$. Red lines refer to empty mass bins.
   }
\label{S_SIG100}
\end{figure}

This restricts the efficient termination of accretion by fragmentation to the
inner parts --e.g. the terrestrial zone in the solar system-- of a planetary
system.  The migration process enters only through the lower cut-off mass
$m_0$. While the uncertainty of $m_0$ in principle is not a big issue, as it
appears in the logarithm, the truth is that the migration timescale depends on
the planetesimal radius which can vary significantly. An uncertainty of the
cut-off radius by a factor of ten indicates an uncertainty of
$M_{\mathrm{mill}}$ of the same order, which again leads us to the question
about the necessity of a careful treatment of migration in a global frame.

All simulations use a lower cut-off size of $800$ metres, which is roughly
equivalent to the {cut-off} introduced by migration. Since $m_0$ is defined by
the identity of the migration timescale and the fragmentation timescale (see
section ``Collisional cascades'' of Paper I), this mass is also independent of
the surface density, given that the ratio of solid to gaseous material is
constant.  While the more refined simulation S6FBH shows a mass loss only
reduced by 30\%, we expect that the uncertainty due to the reduced treatment of
migration to be at least of the same order.

In view of these considerations, we retake now the analysis of the simulations:
Simulation S9\_S100 is strongly affected by the mill process, whereas
simulation S1FB still retains a significant fraction of the initial mass.  The
quiescent conditions in simulation S8\_S2 exclude a prominent role of
fragmentation at any evolutionary stage. Thus we estimate $M_{\mathrm{mill}}
\approx 0.1 M_{\oplus} $ for a solar system analogue at 1 AU (see
figure~\ref{Mmax_SIG}), which yields the approximate expressions:

\begin{align}
 M_{\mathrm{mill}}  & =  f\times 0.1 \, M_{\oplus}\times
  \left(\frac{r}{1 { AU}}\right)^2 \left(\frac{M_c}{1\,M_{\odot}}\right)^{-2/3}\nonumber \\
& \left(\frac{\rho}{2.7\, {g/cm}^3}\right)^{2/3}
  \label{MmillScale}
\end{align}

\noindent
Since the protoplanets maintain a separation of approximately $10\, R_{\mathrm{Hill}}$, the mill mass
corresponds to an upper limit $\Sigma_{\mathrm{mill}}$ of the surface density which is available
for the formation of protoplanets:

\begin{align}
\Sigma_{\mathrm{mill}} &=  \frac{M_{\mathrm{mill}}}{20\pi a R_{\mathrm{Hill}}} \nonumber \\
 &=  f^{2/3}\times 9.15 \, \frac{\mathrm{g}}{\mathrm{cm}^2 }\times
  \left(\frac{r}{1 { AU}}\right)^{-2/3} \left(\frac{M_c}{1\,M_{\odot}}\right)^{-1/9}\nonumber \\
  &\left(\frac{\rho}{2.7\, {g/cm}^3}\right)^{4/9}
\end{align}

The scaling relation~\ref{MmillScale} implies $ M_{\mathrm{mill}} \approx
2.5\,M_{\oplus} $ at 5 AU, which is in agreement with an upper core mass of $4
M_{\oplus} $ found in the simulations of \cite{Ina2003}. Although it seems
impossible to form a core that is large enough ($15\, M_{\oplus}$) to initiate
gas accretion, this tight upper limit is due to disregarding the gaseous
envelope --i.\,e. the protoplanetary atmosphere before the onset of strong gas
accretion-- of the growing core. Since the gaseous envelope enhances the
accretion cross section by an order of magnitude (and hence $f\approx 10$ in
equation~\ref{MmillScale}), the mill mass increases by the same factor. Thus the
formation of a $15\, M_{\oplus}$ proto-jovian core at 5 AU is not ruled out by
fragmentation, again in agreement with \cite{Ina2003}.

Both low--mass simulations S1FB and S8\_S2 still contain a major fraction of
the total mass in the statistical component, which prevents the onset of
orbital crossing on a timescale of a few $10^5$ years. However, the fast
protoplanetary growth in the high--mass simulation S9\_S100, accompanied by an
intense mass loss, leads to an onset of strong protoplanet--protoplanet
interactions already at the end of the simulation.  The chaotic evolution of
the velocity dispersion at the high mass end, as we can see in
figure~\ref{S_SIG100} bottom, indicates an intense interaction of the $N-$body
particles.

\section{Discussion}
\label{sec.discussion}

In this article, which can be envisaged as a continuation of the work we
presented in Paper I, we first carefully assess the code and then apply it to
investigate the formation of protoplanets. Our main results are summarised as follows

\begin{enumerate}

\item The influence of the fragmentation model on the protoplanetary growth is
weak during the fast initial runaway growth. In particular, any realistic
choice of the impact strength does not inhibit the growth of the planetesimals.
However, the choice of the fragmentation model controls the oligarchic growth
through the overall mass loss due to the migration of  smaller fragments. Our
simulations show that the \cite{Housen1990} strength leads to a significant
deceleration of the mass accretion in the later phases. Thus the recent impact
strength from \cite{Benz1999} is more favourable in terms of an efficient
protoplanet formation.

\item We derive the notion of a critical {\em mill mass} to provide a
convenient handle on the fragmentation processes.  If the mass of a
(proto)planet exceeds this critical limit, then an interplay of destructive
collisions and the removal of fragments by migration terminates the accretion
of planetesimals. In particular, this critical mass implies an upper limit of
the mass (in solids), which can be transformed into planets, unless migration
ceases very early due to the fast dissipation of the gaseous disc.

\item Contrary to the work of \cite{Rafikov2001}, we find no termination of the
protoplanetary accretion due to gap formation. None of our simulations shows
any significant radial structure, except for a limited time during the runaway
accretion.  While low surface densities favour gap formation, all observed
radial features are so weak that the notion ``gap'' does not correspond to
these structures.  Hence, resonant interactions between protoplanets and the
field planetesimals are not a dominant process during the growth phases
considered, which also supports the validity of the Fokker--Planck approach.
Likewise, the dynamically hot field planetesimals also suppress
non-axisymmetric features beyond the Hill radius of the protoplanets.  We must
mention that the difference we find with \cite{Rafikov2001} is true for a {\em
hot} disc. Nevertheless, his solution is correct in some cases, especially when the
disc can remain cold \citep[see][]{IdaEtAl2000,KirshEtAl2009} as in, for
instance, the gaps of Saturn's rings \citep[ e.g. the work
of][]{GoldreichTremaine1978,GoldreichTremaine1978b,LissauerEtAl1981}.

\end{enumerate}

The eccentricity and inclination of the protoplanets remain small during the
oligarchic growth phase. However, we note that this does not imply small
eccentricities of the final planets, since the onset of orbital crossing
terminates the dynamically quiet oligarchic growth phase.

Since our work introduced a new computer code to study the growth of
protoplanets, we primarily focussed on the careful assessment of its validity
and a small parameter study to strengthen this approach. Considering that the
current abilities of the hybrid code exclude global simulation which could
address migration in a proper way, we restricted our studies to a small ring of
planetesimals.  However, our experience drawn from this work allows an outline
of possible improvements.  The wallclock time of a rather small simulation is
dominated by the integration of the statistical component. As the radial
extension of the simulation volume is increased, the computing time due to the
statistical component increases linearly, whereas the computing time due to the
{$N-$body} component increases proportional to the square of the radial
width. If the resolution of the radial grid is reduced, the weight of the
{$N-$body} part will further increase. A moderately extended model, which
covers the inner planetary system up to 10 AU, requires the long-term
integration of $10^3$ to $10^4$ particles.

While these are only few particles compared to big star cluster simulations
\citep[e.g.][]{Makino2004,Berczik2005}, the long integration times of at least
$10^6$ orbits prevent the efficient parallelisation. Astrophysicists had an
early start in the field of through the GRAPE hardware in a standard PC cluster
\citep[see the extensive description in][]{Fuku2005}. A more promising solution
are the modern graphics processing units (GPUs), which have made significant
progress in the last years. They were originally used to perform calculations
related to 3D computer graphics.  Nevertheless, due to their highly parallel
structure and computational speed, they can be very efficiently used for
complex algorithms.  Computational astrophysics has been a pioneer to use GPUs
for high performance general purpose computing (see for example the early
AstroGPU workshop in Princeton 2007, through the information
base\footnote{\url{http://www.astrogpu.org}}). The direct $N-$body code has
been ported to GPUs by Sverre Aarseth who, as is his admirable custom, has made
the code publicly available.  We plan on porting our hybrid method to GPU
technology soon.

The extension of the simulations towards longer integration times does not only
require an optimisation of the hybrid code, but also a more careful modelling
of the growing planets to account for the interaction with the gaseous disc.
While these improvements are necessary to allow the consistent treatment of
migration, they also open the study of the early debris disc phase.  Debris
discs could provide constraints on the planet formation process, since the low
opacity of kilometre-sized planetesimals prevents the direct observation of the
protoplanetary growth in extrasolar systems.  Though all these improvements are
not implemented yet, they encourage us to pursue the further development of the
hybrid approach.

\section*{Acknowledgments}

PAS thanks the National Astronomical Observatories of China, the Chinese
Academy of Sciences and the Kavli Institute for Astronomy and Astrophysics in
Beijing, as well as the Pontificia Universidad de Chile and Jorge Cuadra for
his invitation. He shows gratitude to Wenhua Ju, Hong Qi and Xian Chen for
their support and hospitality.  RS acknowledges support by the Chinese Academy
of Sciences Visiting Professorship for Senior International Scientists, Grant
Number 2009S1-5 (The Silk Road Project). The special supercomputer Laohu at the
High Performance Computing Center at National Astronomical Observatories,
funded by Ministry of Finance under the grant ZDYZ2008-2, has been used.
Simulations were also performed on the GRACE supercomputer (grants I/80 041-043
and I/84 678-680 of the Volkswagen Foundation and 823.219-439/30 and /36 of the
Ministry of Science, Research and the Arts of Baden-urttemberg).  Computing
time on the IBM Jump Supercomputer at FZ J{\"u}lich is acknowledged.

\label{lastpage}

\end{document}